\documentclass[aip,pof,preprint]{revtex4-1}
\pdfoutput=1
\usepackage{graphicx}
\usepackage{natbib}
\usepackage{amsmath,amssymb}
\usepackage{subfig}
\usepackage{setspace}

\usepackage[font=small,font=singlespacing,justification=raggedright]{caption}

\usepackage{dcolumn}
\usepackage{hyperref}

\usepackage{ifthen}
\usepackage[usenames, dvipsnames]{color}

\newboolean{showcomments}
\setboolean{showcomments}{true}

\newcommand{\dennice}[1]{  \ifthenelse{\boolean{showcomments}}
{\textcolor{blue}{(Dennice says:  #1)}}{}}
\newcommand{\brian}[1]{\ifthenelse{\boolean{showcomments}}
{\textcolor{Red}{(Brian says: #1)}}{}}
\newcommand{\petros}[1]{\ifthenelse{\boolean{showcomments}}
{\textcolor{Green}{Petros says: #1)}}{}}
\newcommand{\vaughan}[1]{\ifthenelse{\boolean{showcomments}}
{\textcolor{Purple}{(Vaughan says: #1)}}{}}

\begin{document}

\title{A minimal model of self-sustaining turbulence }

\author{Vaughan Thomas}
\affiliation{Department of Mechanical Engineering, Johns Hopkins University, Baltimore, Maryland, 21218, USA}
\author{Brian Farrell}
\affiliation{School of Engineering and Applied Science, Harvard University, Cambridge, Massachusetts, 02138, USA}
\author{Petros Ioannou}
\affiliation{Department of Physics, National and Kapodistrian University of Athens, Panepistimiopolis, Zografos, Athens, 15784, Greece}
\author{Dennice F. Gayme}
\affiliation{Department of Mechanical Engineering, Johns Hopkins University, Baltimore, Maryland, 21218, USA}

\date{\today}

\begin{abstract}

In this work we examine the turbulence maintained in a Restricted Nonlinear (RNL) model of plane Couette flow.  This model is a computationally efficient approximation of the second order statistical state dynamics (SSD) obtained by partitioning the flow into a streamwise averaged mean flow and perturbations about that mean, a closure referred to herein as the RNL$_\infty$ model.  The RNL model investigated here employs a single member of the infinite ensemble that comprises the covariance of the RNL$_\infty$ dynamics.  The RNL system has previously been shown to support self-sustaining turbulence with a mean flow and structural features that are consistent with DNS.  This paper demonstrates that the RNL system's self-sustaining turbulent state is supported by a small number of streamwise varying modes, which form the natural support for the self-sustaining process maintaining RNL turbulence.  Remarkably, truncation of the RNL system's support to a single streamwise varying mode can suffice to sustain the turbulent state. The close correspondence between RNL simulations and DNS that has been previously observed along with the results presented here suggest that the fundamental mechanisms underlying wall-turbulence can be analyzed using these highly simplified RNL systems.

\end{abstract}
\maketitle

\section{Introduction}
The analytical intractability of the Navier Stokes (NS) equations has impeded attempts to develop a comprehensive understanding of the dynamics of turbulence in wall-bounded shear flows.  This impediment to understanding resulting from the complexity of NS dynamics has led to extensive efforts to obtain simplifications of the NS system that still retain fundamental aspects of the dynamics of wall-turbulence.    One useful approach for simplifying the NS system is to study models of reduced order.  Model order reduction can be accomplished by Galerkin projections of the infinite dimensional NS system onto a finite low dimensional space.  The basis functions used are generally chosen  for their  particular properties such as Fourier modes for use in Cartesian channels with diffusive dissipation, Proper Orthogonal Decomposition (POD) projections \cite{Lumley-1967,Smith-etal-2005} for economy in representing the structures occurring in turbulence, and balanced truncation for economical representation of turbulence dynamics. \cite{Farrell-Ioannou-2001-accurate,Rowley2005,Rowley2010}

A second area of research aimed at understanding the dynamics underlying the NS equations is the study of systems in which the complexity of the dynamics has been reduced.  One such method is confining the turbulence to a minimal channel that reduces the complexity of supported perturbations and results in  a simplification of the flow structures.  This method was used to demonstrate the importance of large-scale roll and streak structures in maintaining turbulence. In particular, a minimal channel was used to show that  wall-turbulence does not self-sustain unless  the confining channel is large enough to accommodate roll and streak structures of sufficient spanwise  and  streamwise extent. \cite{Jimenez-Moin-1991,Hamilton-etal-1995}

Reduction of the support of turbulence has also been attempted by seeking a skeleton of exact coherent structures in the phase space of the transient turbulent attractor, see e.g. \cite{Jimenez-etal-2005,Kawahara-etal-2012}.   For plane Couette flow, the first such  unstable solution was the fixed point computed by \citet{Nagata-1990}. Subsequent research using numerical methods has uncovered additional fixed points and periodic orbits for plane Couette flow. \cite{Gibson-etal-2009,Kawahara-etal-2012}   While promising conceptually, the project of extending these solutions to the global turbulent dynamics has yet to be completed. A related approach is to isolate and model specific flow features and interactions in a schematic fashion, examples of this approach include the roll/streak instability SSP model of  Waleffe~\cite{Waleffe-1995a}, the nine-mode truncation model of Moehlis et al. \cite{Moehlis-etal-2004},  and the model describing laminar-turbulent spots and structures by \citet{Tuckerman2011}.

Another method of reducing the complexity of turbulence dynamics is to simplify the equations themselves. One example of this approach is to study turbulence using  the Linearized Navier Stokes (LNS) equations, which can  be analyzed comprehensively using  linear systems theory. \cite{Farrell-Ioannou-1996a,Farrell-Ioannou-1996b} The LNS equations capture a number of fundamental aspects of turbulence dynamics including the non-normal disturbance amplification mechanism
\cite{Farrell-1988a,Gustavsson-1991,Trefethen-etal-1993,Reddy-Henningson-1993,Farrell-Ioannou-1993e,Bamieh-Dahleh-2001,Jovanovic-Bamieh-2005,Butler-Farrell-1992}, which has been shown to be necessary for both energy production in fully turbulent flow \cite{Henningson-1996b} and to play a key role in the bypass transition mechanism.~\cite{Henningson-Reddy-1994} The LNS equations also provide specific insights into the mechanism maintaining turbulence including the role of the coupling between the Orr-Sommerfeld and Squire equations in generating the robust transient growth of streaks by the lift-up process that is  an integral component of the roll/streak mechanism underlying the self-sustaining process (SSP) in wall-bounded shear flows.\cite{Butler-Farrell-1993,Kim-Lim-2000}  The LNS equations have also proven successful in predicting  second-order statistics in these flows. \cite{jovbamCDC01,Zare-etal-2014a,Zare-etal-2015,Butler-Farrell-1993,Farrell-Ioannou-1998a,DelAlamo-Jimenez-2006,Hwang-Cossu-2010,Cossu-etal-2009,moajovJFM12}

The current work employs a statistical state dynamics (SSD) model that incorporates a combination of the reduction of order and the simplification of dynamics approaches.    This  model is based on the restricted nonlinear (RNL) dynamical system, which is a second order closure of the dynamics of the turbulent statistical state comprising the joint evolution of the streamwise constant mean flow (first cumulant) and  the ensemble  second order perturbation statistics (second  cumulant).     The SSD is closed either by parametrizing the higher cumulants using  stochastic excitation \cite{Farrell-Ioannou-1993e,DelSole-Farrell-1996,DelSole-04} or by setting the third cumulant to zero, see e.g \cite{Marston-etal-2008,Tobias-etal-2011,Srinivasan-Young-2012}.  Restricting the Navier-Stokes (NS) equations  to the first two cumulants retains the nonlinear interaction between the perturbation momentum fluxes and the mean flow but neglects explicit calculation of the nonlinear interactions among the streamwise varying perturbations.  Turbulence was first studied in an RNL modeling framework using the Stochastic Structural Stability Theory system~\cite{Farrell-Ioannou-2003-structural}, which we will refer to as the RNL$_\infty$ system. In the RNL$_\infty$ system  the second cumulant is obtained  by solving the time dependent Lyapunov equation for the perturbation covariance  dynamics.  This time dependent Lyapunov equation represents the dynamics of an infinite ensemble of realizations of the perturbation structure.  As a result, the SSD of the RNL$_\infty$ system is autonomous, which is particularly useful for obtaining analytical results.  However, the covariance matrix dimension for a perturbation dynamics of $O(N)$  is $O(N^2)$, which limits the resolution of models that can be studied directly using the RNL$_\infty$ model.   In order to overcome this limitation in subsequent implementations of the RNL model  the covariance was estimated from a single realization of the perturbation dynamics.~\cite{Thomas2014,Constantinou-etal-2014}  This approximation shares the dynamical restrictions of the RNL$_\infty$ system but unlike the RNL$_\infty$ model it retains small fluctuations in the perturbation covariance due to the perturbation covariance dynamics not being that of an infinite ensemble. It is this dynamical restriction that is primarily responsible for the insights into turbulence dynamics obtained from the RNL framework.\cite{Constantinou-etal-2014,Thomas2014} The existence of fluctuations in the approximate covariance does not greatly affect the correspondence between simulation results obtained using the analytically tractable RNL$_\infty$ and those obtained with the numerically tractable RNL implementation used in this work.\cite{Thomas2014}

The goals of this work are to further investigate the dynamics of RNL turbulence and to examine the implications of its simplified structure for understanding wall-turbulence.  The primary focus of our study is the mechanisms that maintain the turbulent flow and determine the structure of its statistical mean state.   It is natural that in studying these mechanisms attention is focused on the dynamics of the streak and roll structures. These structures, which are ubiquitous in wall-turbulence, respectively comprise the large-scale streamwise streaks of high and low speed fluid and the associated streamwise vortices the circulations of which reinforce the streaks through the linear non-normal  lift-up mechanism.  Understanding the dynamical mechanisms determining the structure of the mean and perturbation fields and maintaining the statistical equilibrium state of wall-turbulence requires understanding the dynamics of these structures.   The rolls and streaks in wall-turbulence are not associated with an unstable linear mode but rather, are maintained by a nonlinear instability process that is associated with linear non-normal growth of both the streamwise invariant roll and streak structures and the streamwise varying perturbation field that maintains the roll structure. A dynamical mechanism advanced to explain this  nonlinear instability is referred to as a self-sustaining process (SSP).  The first such process traced the origin of the perturbations required to sustain  linear non-normal streak growth  to the  break-up of previously generated streaks.\cite{Swearingen-Blackwelder-1987,Blakewell-Lumley-1967, Jimenez-2013,Schoppa-Hussain-2002} An alternative explanation of the  SSP suggested that Reynolds stresses arising from inflectional instability of the streak maintain the roll circulations. \cite{Hamilton-etal-1995,Waleffe-1997,Hall-Sherwin-2010}  However, analysis of simulations subsequently revealed that most streaks are too weak to sustain strong inflectional instability, which resulted in the suggestion that transient growth gives rise to the  roll-maintaining perturbations. \cite{Schoppa-Hussain-2002} Transiently growing perturbations have also been shown to maintain the  roll circulation in the RNL system. However, in contrast to previously proposed transient  growth based mechanisms  \cite{Jimenez-2013,Schoppa-Hussain-2002}, the transiently growing perturbations in RNL turbulence result not from random the occurrence of optimal perturbations associated with streak breakdown but rather from systematic parametric instability of the time-dependent streak. \cite{Farrell-Ioannou-1996b, Farrell-Ioannou-1999, Farrell-Ioannou-2012}

In this work we study the RNL dynamics both by approximating the closure for the third cumulant of the RNL$_\infty$ SSD using a stochastic forcing and by approximating the closure of this SSD that sets the third cumulant to zero.  The main results demonstrate that RNL turbulence is naturally maintained solely through interactions between the  streamwise mean flow (the $k_x=0$ streamwise Fourier components) and a small set of streamwise varying modes (i.e. the $k_x\neq 0$ streamwise Fourier components).  Moreover, we establish that a minimal configuration for maintaining turbulence can be obtained by limiting the streamwise varying wavenumber support of the RNL turbulence to  a single streamwise varying mode. A second contribution of this work is an investigation of RNL turbulence in extended channels. Specifically, we show that RNL turbulence self-sustains in channels with streamwise extents of $96\pi\delta$, where $\delta$ is the half channel height. These results suggest that self-sustaining RNL turbulence continues to exist in channels with infinite streamwise extent.

This paper is organized as follows. The next section contains a derivation of  the RNL model from the NS equations and establishes its relation to the RNL$_\infty$ system. In Section \ref{sec:numerics} we describe our numerical approach. Then in Section \ref{sec:results} we show that the RNL system naturally maintains a turbulent state that is consistent with that of DNS and that RNL turbulence is supported by only a small set of streamwise modes. In Section \ref{sec:tuncatingRNLsupport}  we truncate the streamwise varying wavenumber support of the RNL dynamics to a single mode and explore the interval in streamwise wavenumber over which the turbulence is sustained. We also explore the sensitivity of the turbulence statistics to the retained wavenumber. Finally, we conclude the paper and point to directions of future study.

\section{Modeling framework}
\label{sec:framework}

 Consider a plane Couette flow between walls with velocities $\pm U_w $. The streamwise direction is $x$, the wall-normal direction is $y$, and the spanwise direction is $z$. Quantities  are non-dimensionalized by the channel half-width, $\delta$, and the wall velocity, $U_w$. The lengths of the channel in the streamwise and spanwise  directions are respectively $ L_x$ and $L_z $. Streamwise averaged, spanwise averaged, and time-averaged quantities are respectively denoted  by angled brackets, $\langle\,\bullet\,\rangle = \tfrac{1}{L_x} \int_0^{ L_x} \bullet ~  {\rm{d}} x $, square brackets, $[\bullet]=\tfrac{1}{L_z} \int_0^{L_z} \bullet ~{\rm{d}} z$, and an overline $\overline{\;\bullet\;} = \tfrac{1}{T} \int_0^T \bullet~{\rm{d}} t$, with  $T$ sufficiently large. The velocity field $\mathbf{u}_T$ is decomposed into a streamwise averaged mean, $\mathbf{U}(y,z,t)=(U,V,W)$, and the deviation from this mean (the perturbation), $\mathbf{u}(x,y,z,t)=(u,v,w)$.  The pressure gradient is similarly decomposed  into a streamwise averaged mean, $\nabla P(y,z,t)$, and the deviation from this mean, $\nabla p(x,y,z,t) $. The corresponding Navier Stokes (NS) equations are
 \begin{subequations}
 \allowdisplaybreaks
\label{eq:NSE0}
\begin{eqnarray}
&&\mathbf{U}_t+ \mathbf{U} \cdot \nabla \mathbf{U} + \nabla P - \frac{1}{R}\Delta \mathbf{U} = - \langle\mathbf{u} \cdot \nabla \mathbf{u}\rangle,
\label{eq:NSm}\\
&& \mathbf{u}_t+   \mathbf{U} \cdot \nabla \mathbf{u} +
\mathbf{u} \cdot \nabla \mathbf{U}  + \nabla p -  \frac{1}{R}\Delta  \mathbf{u}
= -  \left(\mathbf{u} \cdot \nabla \mathbf{u} - \langle\mathbf{u} \cdot \nabla \mathbf{u}\rangle \right)  + \epsilon
 \label{eq:NSp}\\
&&\nabla \cdot \mathbf{U} = 0,~~~\nabla \cdot \mathbf{u} = 0 , \label{eq:1c}~~~
\end{eqnarray}
\end{subequations}
 where $\varepsilon$ is a stochastic excitation used to initiate turbulence and the Reynolds number is defined as $R = {U_w \delta}/{\nu}$, with kinematic viscosity $\nu$.

 Obtaining the statistical state dynamics (SSD) corresponding to
\eqref{eq:NSE0} requires solving the
infinite hierarchy of cumulant equations, see e.g. \cite{Frisch-1995,Hopf-1952}. However,  a useful and tractable
approximation to the full SSD is obtained by closing this infinite hierarchy  at second order
by either neglecting the third cumulant or parametrizing it appropriately. We refer to this approximation as
the RNL$_\infty$ system and use the $\infty$  to indicate that the
ensemble of realizations is infinite. Physical realizations of the RNL$_\infty$ system can be obtained by making
the ergodic assumption that the ensemble averages and streamwise averages are equal  and consequently the first cumulant
is the streamwise averaged flow. In this work we will approximate the second cumulant of the RNL$_\infty$ dynamics
by the streamwise average of the spatial two-point correlations
obtained from a single $\mathbf{u}$ field realization.
The resulting approximation to the RNL$_\infty$ system is governed
by the following equations:
 \begin{subequations}
 \allowdisplaybreaks
\label{eqn:RNL}
\begin{eqnarray}
&&{ \mathbf{U}_t}+   \mathbf{U} \cdot \nabla \mathbf{U}
 + \nabla  P - \frac{1}{R} \Delta \mathbf{U}  = -
 \langle\mathbf{u} \cdot \nabla \mathbf{u}\rangle,
\label{eqn:RNL-mean}\\
&&{ \mathbf{u}_t}+   \mathbf{U} \cdot \nabla \mathbf{u} +
\mathbf{u} \cdot \nabla \mathbf{ U}  + \nabla  p -  \frac{1}{R} \Delta \mathbf{u} = \mathbf{e} \label{eqn:RNL-perturb1}\\
 &&\nabla \cdot \mathbf{U} = 0,~~~\nabla \cdot \mathbf{u} = 0 .~~~\label{eqn:RNL-div}
 \end{eqnarray}
\end{subequations}
We refer to equation \eqref{eqn:RNL} as the RNL model. The dynamics in \eqref{eqn:RNL} correspond to parametrizing the  perturbation-perturbation nonlinearity, $\mathbf{u} \cdot \nabla \mathbf{u} - \langle\mathbf{u} \cdot \nabla \mathbf{u}\rangle$ in \eqref{eq:NSp}, with the stochastic excitation $\mathbf{e}$. These equations  approximate the closure of the SSD at second order using a stochastic parametrization of the higher order cumulants or alternatively by neglecting the third cumulant, which corresponds to setting  $\mathbf{e}$ = $\mathbf{0}$.
The nonlinear equation  \eqref{eqn:RNL-mean} describes the dynamics of a streamwise averaged mean flow driven by the divergence of the streamwise averaged Reynolds stresses, which we denote by, e.g. $\langle{uu}\rangle$, $\langle{uv}\rangle$.
Equation \eqref{eqn:RNL-perturb1} describes the influence of  the  streamwise constant mean  flow, $\mathbf{U}(y,z,t)$, on the linearized perturbation dynamics.

\section{Numerical approach }
\label{sec:numerics}

The numerical simulations in this paper were carried out using a spectral code based on the Channelflow NS equations solver.\cite{channelflow,Gibson-etal-2008} The time integration uses a third order multistep semi-implicit Adams-Bashforth/backward-differentiation scheme that is detailed in \cite{Peyret2002}. The discretization time step is automatically adjusted such that the Courant-Friedrichs-Lewy (CFL) number is kept between $0.05$ and $0.2$. The spatial derivatives employ Chebyshev polynomials in the wall-normal ($y$) direction and Fourier series expansions in the streamwise ($x$) and spanwise ($z$) directions. \citep{Canuto1988} No-slip boundary conditions are employed at the walls and periodic boundary conditions are used in the $x$ and $z$ directions. Aliasing errors from the Fourier transforms are removed using the 3/2-rule detailed in \citet{Zang1985}. A zero pressure gradient is imposed in all simulations. Table \ref{table:geometry} provides the dimensions of the computational box, the number of grid points, and the number of spectral modes for the DNS and RNL simulations. In both the DNS  and RNL simulations we use the respective stochastic excitations $\epsilon$ in \eqref{eq:NSE0}  and $\mathbf{e}$ in \eqref{eqn:RNL} only to initiate turbulence. In order to perform the RNL computations the DNS code was restricted to the dynamics of \eqref{eqn:RNL}.

\begin{table*}[!t]
\centering
\caption{\label{table:geometry}Geometry for the numerical simulations. $x/\delta$, $y/\delta$ and $z/\delta$ define the computational domain, non-dimensionalized by the channel half-height, $\delta$. $N_x$, $N_y$ and $N_z$ are the number of grid points in their respective directions. $M_x$ and $M_z$ are the number of Fourier modes used before dealiasing and $M_y$ is the number of Chebyshev modes used in each simulation.}
\begin{tabular}{p{2.cm}p{1.5cm}p{1.5cm}p{1.5cm}p{3cm}p{2.5cm}} \hline
 Case & \parbox[c]{1cm}{$x/\delta$} &\parbox[c]{1cm}{$y/\delta$} &\parbox[c]{1cm}{$z/\delta$} & $N_x \times N_y \times N_z$ & $M_x \times M_y \times M_z$  \\ \hline
DNS-A  & $[0, \ 4\pi]$ & $[-1,1]$ & $[0,4\pi]$ & $128 \times 65 \times 128$  & $128\times 65 \times 65$ \\
DNS-B  & $[0, \ 8\pi]$ & $[-1,1]$ & $[0,4\pi]$ & $128 \times 65 \times 128$  & $128\times 65 \times 65$ \\
DNS-C  & $[0, \ 12\pi]$ & $[-1,1]$ & $[0,4\pi]$ & $128 \times 65 \times 128$  & $128\times 65 \times 65$ \\
DNS-D  & $[0, \ 16\pi]$ & $[-1,1]$ & $[0,4\pi]$ & $128 \times 65 \times 128$  & $128\times 65 \times 65$ \\
RNL-A  & $[0, \ 4\pi]$ & $[-1,1]$ & $[0,4\pi]$ & $\;\,16 \times 65 \times 128$ & $\;\,16\times 65 \times 65$ \\
RNL-B  & $[0, \ 8\pi]$ & $[-1,1]$ & $[0,4\pi]$ & $\;\,32 \times 65 \times 128$ & $\;\,32\times 65 \times 65$ \\
RNL-C  & $[0,12\pi]$ & $[-1,1]$ & $[0,4\pi]$ & $\;\,48 \times 65 \times 128$ & $\;\,48\times 65 \times 65$ \\
RNL-D  & $[0,16\pi]$ & $[-1,1]$ & $[0,4\pi]$ & $\;\,64 \times 65 \times 128$ & $\;\,64\times 65 \times 65$ \\
RNL-E  & $[0,24\pi]$ & $[-1,1]$ & $[0,4\pi]$ & $\;\,96 \times 65 \times 128$ & $\;\,96\times 65 \times 65$ \\
RNL-F  & $[0,32\pi]$ & $[-1,1]$ & $[0,4\pi]$ & $\;\,96 \times 65 \times 128$ & $\;\,96\times 65 \times 65$ \\
RNL-G  & $[0,48\pi]$ & $[-1,1]$ & $[0,4\pi]$ & $\;\,96 \times 65 \times 128$ & $\;\,96\times 65 \times 65$ \\
RNL-H & $[0,72\pi]$ & $[-1,1]$ & $[0,4\pi]$ & $138 \times 65 \times 128$ & $138\times 65 \times 65$ \\ \hline
\end{tabular}
\end{table*}

We also perform a number of flow simulations where the flow dynamics of the RNL model are restricted to a single streamwise varying perturbation and the streamwise averaged mean flow. We implement these cases by introducing a damping term, $\xi\mathbf{u}$, into the equation for the streamwise varying perturbations \eqref{eqn:RNL-perturb1} in the following manner.
\begin{equation}
{ \mathbf{u}_t}+   \mathbf{U} \cdot \nabla \mathbf{u} +
\mathbf{u} \cdot \nabla \mathbf{ U}  + \nabla  p -  \frac{1}{R} \Delta \mathbf{u}  -\xi \mathbf{u}  = \mathbf{e} \label{eqn:RNL-perturb_trun}
\end{equation}
The geometric parameters for this set of cases is given in Table \ref{table:kxTruncationGeometry} and a description of their implementation is provided in Section \ref{sec:tuncatingRNLsupport}.
\begin{table*}[!tb]
\centering
\caption{\label{table:kxTruncationGeometry}Geometry for the numerical simulations of the truncated RNL system. $\lambda_d$ is the wavelength of the undamped streamwise varying perturbations. $x/\delta$, $y/\delta$ and $z/\delta$ define the computational domain, non-dimensionalized by the channel half-height, $\delta$. $N_x$, $N_y$ and $N_z$ are the number of grid points in their respective directions. $M_x$ and $M_z$ are the number of Fourier modes used before dealiasing and $M_y$ is the number of Chebyshev modes used in each simulation.}
\begin{tabular}{p{1.3cm}p{1.9cm}p{1.8cm}p{1.5cm}p{1.5cm}p{3cm}p{2.5cm}} \hline
Case &\parbox[c]{1cm}{$\lambda_d$} &\parbox[c]{1cm}{$x/\delta$} &\parbox[c]{1cm}{$y/\delta$} &\parbox[c]{1cm}{$z/\delta$} & $N_x \times N_y \times N_z$ & $M_x \times M_y \times M_z$  \\ \hline
D1 & $4\pi\delta$ & $[0, \ 4\pi]$ & $[-1,1]$ & $[0,4\pi]$ & $\;\,32 \times 65 \times 128$  & $\;\,32\times 65 \times 65$ \\
D2 &$2\pi\delta$  & $[0, \ 4\pi]$ & $[-1,1]$ & $[0,4\pi]$ & $\;\,32 \times 65 \times 128$  & $\;\,32\times 65 \times 65$ \\
D3  & $4\pi\delta/3$  & $[0, \ 4\pi]$ & $[-1,1]$ & $[0,4\pi]$ & $\;\,32 \times 65 \times 128$  & $\;\,32\times 65 \times 65$ \\
D4  & $\pi\delta$  & $[0, \ 4\pi]$ & $[-1,1]$ & $[0,4\pi]$ & $\;\,32 \times 65 \times 128$  & $\;\,32\times 65 \times 65$ \\
D5  & $4\pi\delta/5$ & $[0, \ 4\pi]$ & $[-1,1]$ & $[0,4\pi]$ & $\;\,32 \times 65 \times 128$  & $\;\,32\times 65 \times 65$ \\
D6  & $2\pi\delta/3$& $[0, \ 4\pi]$ & $[-1,1]$ & $[0,4\pi]$ & $\;\,32 \times 65 \times 128$  & $\;\,32\times 65 \times 65$ \\
D7   &$4\pi\delta/7$ & $[0, \ 4\pi]$ & $[-1,1]$ & $[0,4\pi]$ & $\;\,32 \times 65 \times 128$  & $\;\,32\times 65 \times 65$ \\
K2   & $24\pi\delta$ & $[0, \ 48\pi]$ & $[-1,1]$ & $[0,4\pi]$ & $\;\,96 \times 65 \times 128$  & $\;\,96\times 65 \times 65$ \\
K3   & $16\pi\delta$ & $[0, \ 48\pi]$ & $[-1,1]$ & $[0,4\pi]$ & $\;\,96 \times 65 \times 128$  & $\;\,96\times 65 \times 65$ \\
K4   &$12\pi\delta$& $[0, \ 48\pi]$ & $[-1,1]$ & $[0,4\pi]$ & $\;\,96 \times 65 \times 128$  & $\;\,96\times 65 \times 65$ \\
K5   & $48\pi\delta$/5 & $[0, \ 48\pi]$ & $[-1,1]$ & $[0,4\pi]$ & $\;\,96 \times 65 \times 128$  & $\;\,96\times 65 \times 65$ \\
K12  &$4\pi\delta$ & $[0, \ 48\pi]$ & $[-1,1]$ & $[0,4\pi]$ & $\;\,96 \times 65 \times 128$  & $\;\,96\times 65 \times 65$ \\
K16  & $3\pi\delta$ & $[0, \ 48\pi]$ & $[-1,1]$ & $[0,4\pi]$ & $\;\,96 \times 65 \times 128$  & $\;\,96\times 65 \times 65$ \\
K20 & $12\pi\delta/5$ &$[0, \ 48\pi]$ & $[-1,1]$ & $[0,4\pi]$ & $\;\,96 \times 65 \times 128$  & $\;\,96\times 65 \times 65$ \\
K24 &$2\pi\delta$ &$[0, \ 48\pi]$ & $[-1,1]$ & $[0,4\pi]$ & $\;\,96 \times 65 \times 128$  & $\;\,96\times 65 \times 65$ \\
K30 &$8\pi\delta$/5 &$[0, \ 48\pi]$ & $[-1,1]$ & $[0,4\pi]$ & $\;\,96 \times 65 \times 128$  & $\;\,96\times 65 \times 65$ \\
K40 & $6\pi\delta$/5 &$[0, \ 48\pi]$ & $[-1,1]$ & $[0,4\pi]$ & $128 \times 65 \times 128$  & $128\times 65 \times 65$ \\
K50 & $24\pi\delta$/25 &$[0, \ 48\pi]$ & $[-1,1]$ & $[0,4\pi]$ & $160 \times 65 \times 128$  & $128\times 65 \times 65$ \\
K60 & $4\pi\delta$/5 &$[0, \ 48\pi]$ & $[-1,1]$ & $[0,4\pi]$ & $160 \times 65 \times 128$  & $128\times 65 \times 65$ \\\hline
\end{tabular}
\end{table*}

\section{Results}
\label{sec:results}
In this section we first demonstrate the agreement of RNL simulations with DNS in different channel configurations. We then show that turbulence in this RNL system is naturally supported by a small number of streamwise varying modes. In addition,  the system can be truncated so that it is supported by a single streamwise varying mode interacting with the mean flow. All simulations in this section are at $R=1000$; the full parameter set for each simulation is given in tables \ref{table:geometry} and \ref{table:kxTruncationGeometry}.

\begin{figure}[h]
\subfloat[\label{fig:TurbulentProfile}]{
\includegraphics[width = 0.45\textwidth,clip=]{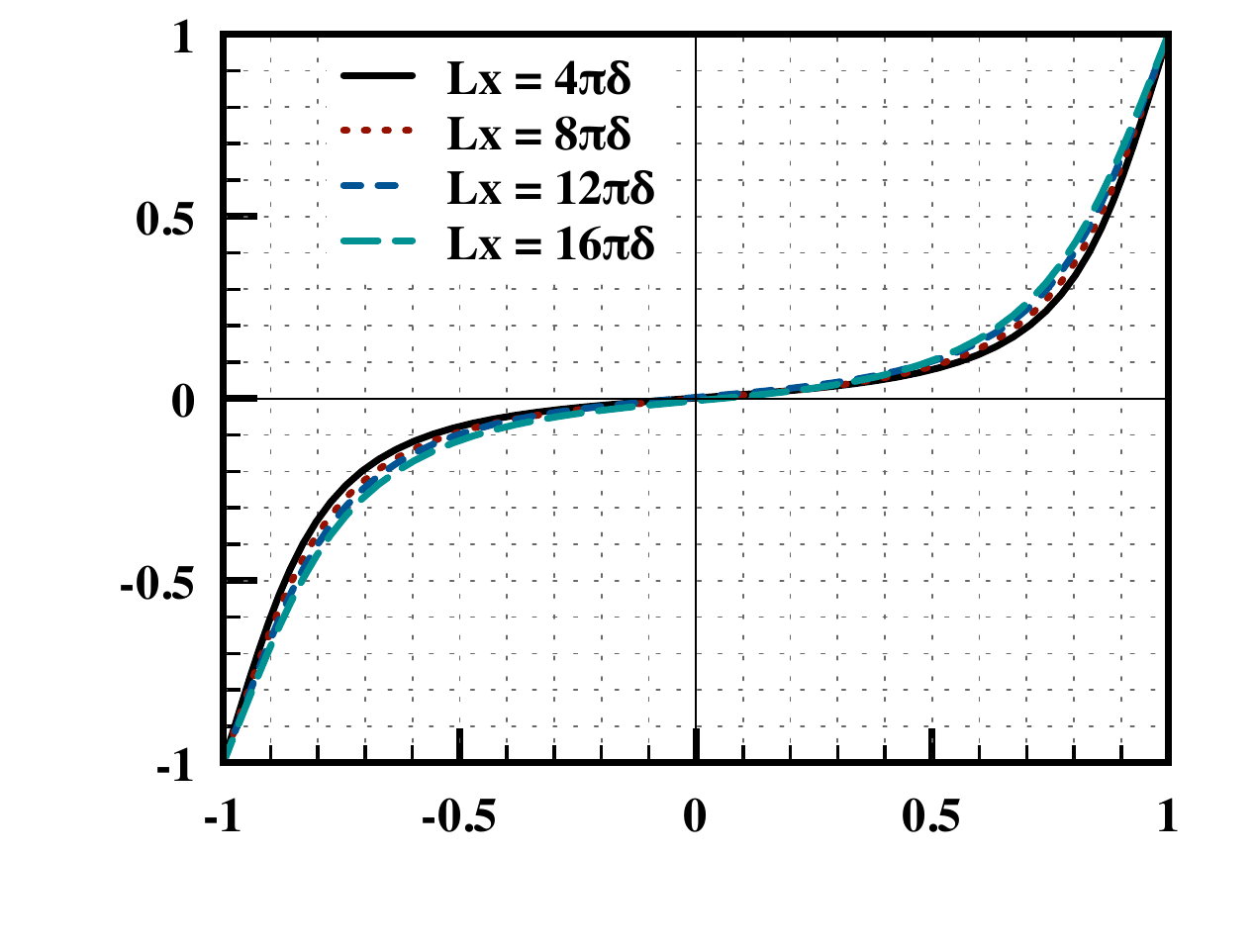}}
\subfloat[\label{fig:WallUnitTurbulentProfile}]{
\includegraphics[width = 0.45\textwidth,clip=]{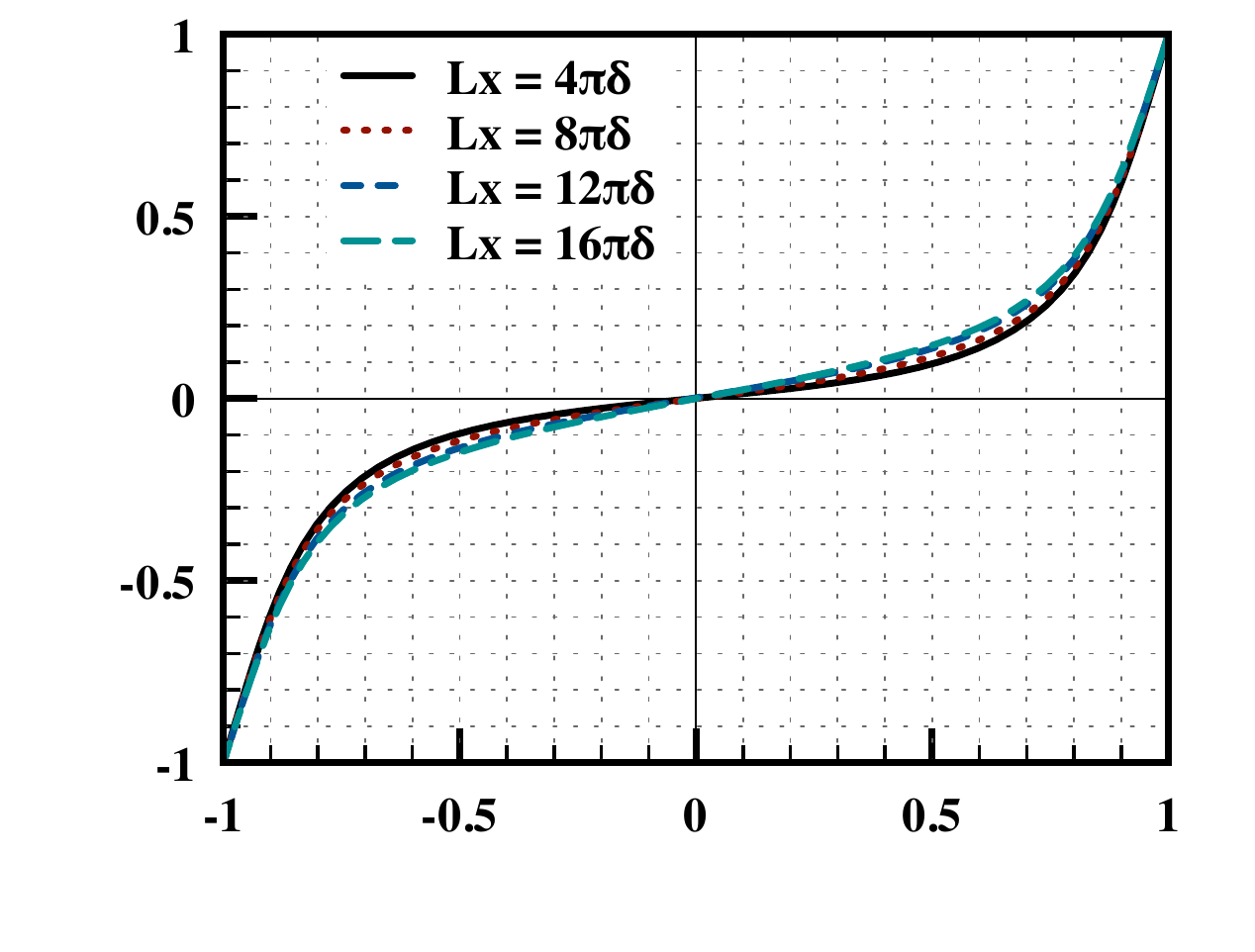}}
\caption{Turbulent mean velocity profiles (based on a streamwise, spanwise and time average) from (a) simulations of the RNL model and (b) DNS for channels with $L_x = \{4\pi, 8\pi, 12\pi,16\pi \}\delta$. }
\label{fig:turbulentprofiles}
\end{figure}

\begin{figure}[t]
\subfloat[\label{fig:RNL_support}]{
\includegraphics[width = 0.45\textwidth,clip=]{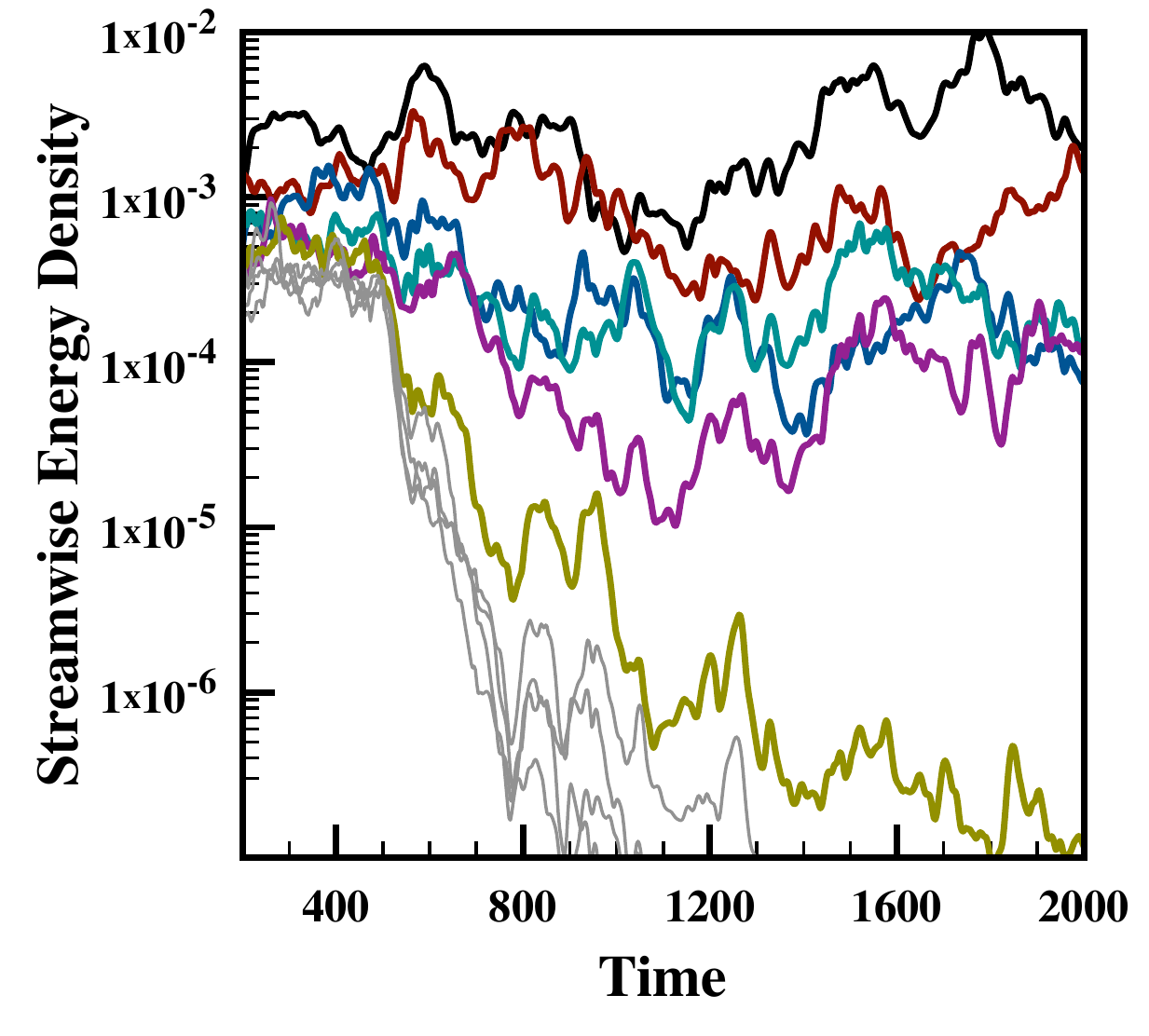}}
\subfloat[\label{fig:DNS_support}]{
\includegraphics[width = 0.45\textwidth,clip=]{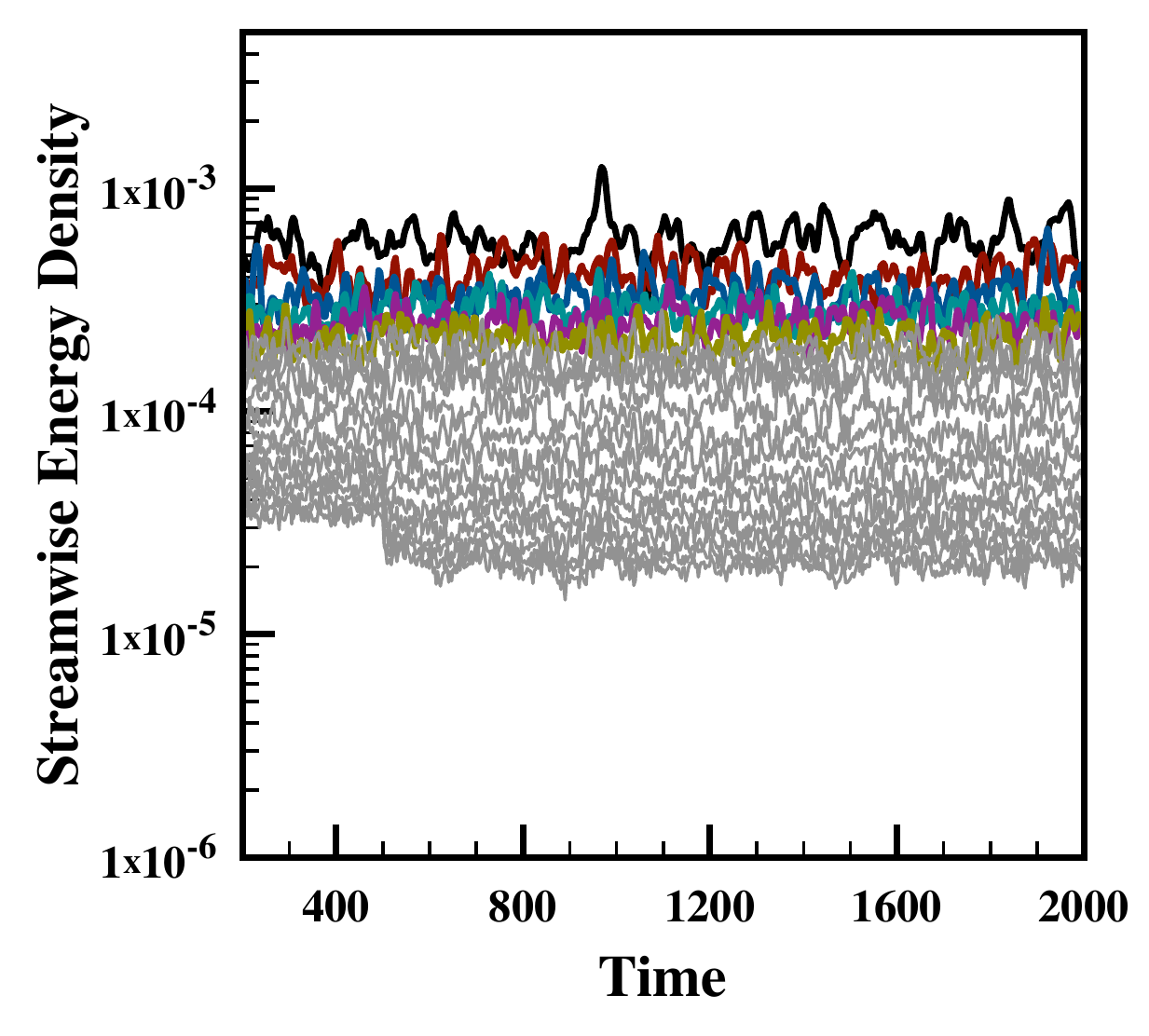}}
\caption{Selected streamwise energy densities for cases (a) RNL-D and (b) DNS-D at $R$ = 1000 for a channel with $L_x$ = 16$\pi\delta$. The energy densities of the streamwise varying perturbations that are supported in the RNL simulation are shown in the following manner $\lambda_1$ = 16$\pi \delta$(black), $\lambda_2$ = 8$\pi \delta$(red), $\lambda_3$ = 16$\pi \delta /3$(blue), $\lambda_4$ = 4$\pi \delta$(cyan), $\lambda_5$ = 16$\pi \delta /5$(violet), $\lambda_6$ = 8$\pi \delta /3$(gold).  The modes that decay when the RNL is in a self-sustaining state are shown in grey in both panels.}
  \label{fig:natural_support}
 \end{figure}

\subsection{The natural support of the RNL system}
\label{sec:RNLnaturalsupport}
 RNL system has been shown to maintain turbulence with a mean flow and structural features that closely resemble those of DNS.\cite{Farrell-Ioannou-2012,Constantinou-etal-2014, Thomas2014} In the current work we study the RNL system described by \eqref{eqn:RNL} across a range of channel configurations.  Figure \ref{fig:turbulentprofiles} demonstrates that RNL simulations produce mean velocity profiles consistent with those from DNS in channels with streamwise extents varying from $4\pi\delta$ to $16\pi\delta$.  Based on the results shown in Figure \ref{fig:turbulentprofiles} and observations that the flow structures are consistent with those reported in \cite{Thomas2014}, we conclude that the RNL system captures the essential features of turbulent flow over a range of operating conditions.  In addition, we observe that varying the channel length does not significantly affect the nature of turbulence sustained in the RNL system.
    We now show that RNL turbulence is naturally supported by a greatly reduced set of streamwise varying wavenumbers. In particular, we demonstrate that when $\mathbf{e}$ in equation \eqref{eqn:RNL-perturb1} is set to 0, the RNL model reduces to a minimal configuration in which only a finite number of streamwise varying perturbations are maintained while energy in the other streamwise varying perturbations decays exponentially. We refer to the set of streamwise wavenumbers that remain when $\mathbf{e}=0$ in equation \eqref{eqn:RNL-perturb1}
as the  natural support for the RNL system.
    In order to quantify the energy in the perturbations we define the streamwise energy density as the perturbation energy associated with streamwise wavelength $\lambda_n$, which is given by
    \begin{equation}
    E_{\lambda_n}(t) = \int_0^{L_z} \int_{-\delta}^{\delta} \int_0^{L_x} \frac{1}{2} \mathbf{u}_{\lambda_n}(x,y,z,t)^2 \ dx \ dy \ dz.
    \end{equation}
Here $\mathbf{u}_{\lambda_n}$ is the perturbation, $\mathbf{u}=(u,v,w)$, associated with Fourier components with streamwise wavelength $\lambda_n$. 

    Figure \ref{fig:natural_support} compares the natural support of the RNL system to the full streamwise energy spectrum supporting turbulence in DNS. Figures \ref{fig:RNL_support} and \ref{fig:DNS_support} shows the time evolution of the streamwise energy densities, $E_{\lambda_{n}}$, for cases  RNL-D and DNS-D, respectively.
Both of these simulations were initiated with a stochastic excitation, respectively $\epsilon$ in \eqref{eq:NSp} and $\mathbf{e}$ in \eqref{eqn:RNL-perturb1}, containing a full range of streamwise and spanwise Fourier components. The excitation was terminated at $t=500$. Figure \ref{fig:DNS_support} demonstrates that in the DNS all streamwise perturbations remain supported. In contrast, the number of streamwise Fourier components supporting RNL turbulence rapidly converges to the small number in the natural set. For the RNL simulation in a 16$\pi\delta$ channel the energy densities associated with all streamwise wavelengths $\lambda_x \leq 8/3\pi\delta$ decay rapidly after the stochastic excitation is removed. These results demonstrate that simulations using the RNL system require substantially less computational resources than  DNS. In what follows, we will further characterize the natural support of turbulence in the RNL system by examining the streamwise energy density as a function of the streamwise extent of the channel.

The difference between the unforced RNL system and the NS equations is the removal of the perturbation-perturbation nonlinearity term, $ \mathbf{u} \cdot \nabla \mathbf{u} - \langle\mathbf{u} \cdot \nabla \mathbf{u}\rangle$. The emergence of the natural set of streamwise varying modes that support self-sustaining RNL turbulence demonstrates the existence of a set of active perturbations that are maintained though interactions with the time-dependent streamwise averaged mean flow. The perturbations that naturally decay in the unforced RNL system cannot be sustained without excitation from nonlinear perturbation-perturbation interactions. The fact that the absence of these weakly interacting perturbations does not significantly alter either the mean flow or the maintenance of turbulence suggests that these perturbations do not play a significant role either in the maintenance or regulation of the RNL turbulent state and are in this sense inactive. Previous work showing the close correspondence in the mean profile and the time-averaged Reynolds stress components shown of Figures 1 and 6 in \citet{Thomas2014}  underscores this point.

The natural support of RNL turbulence in a channel with a streamwise extent of $16\pi\delta$  consists of all of the streamwise varying modes with wavelengths longer than $8/3\pi\delta$.  This lower limit was previously observed to be $4\pi/3\delta$ for case RNL-A. Each of the RNL simulations described in Table \ref{table:geometry} demonstrate a similar lower limit on the wavelengths included in their natural support.  Based on the plot in Figure \ref{fig:RNL_support} it remains unclear whether there is an upper bound on the wavelengths comprising the natural set. In order to investigate this question further we next consider the unforced RNL model in longer channels.

\begin{figure}
\subfloat[]{\label{fig:32PRNLSpectra}
\includegraphics[width = 0.45\textwidth,clip=]{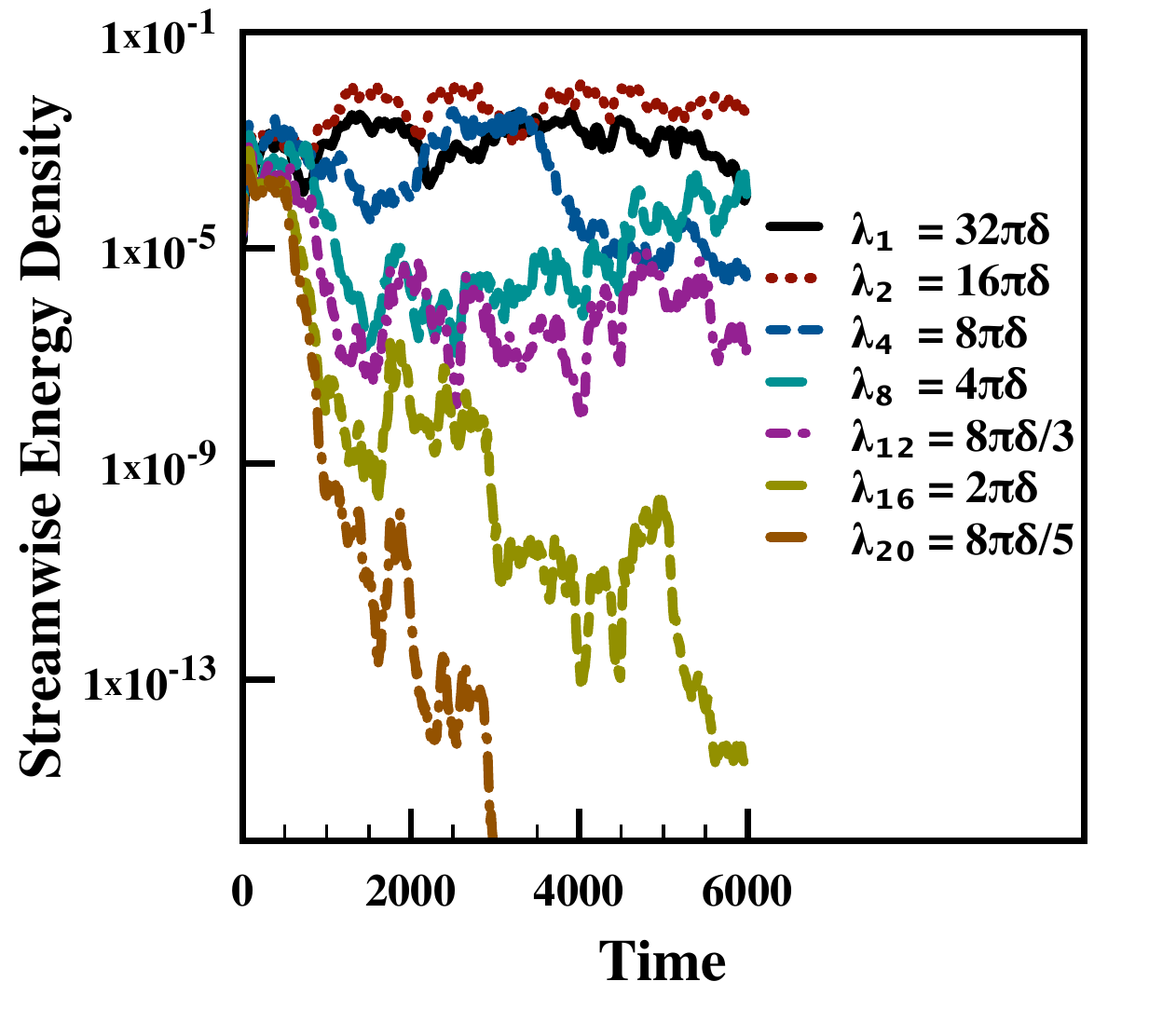} }
\subfloat[]{\label{fig:48PRNLSpectra}
\includegraphics[width = 0.45\textwidth,clip=]{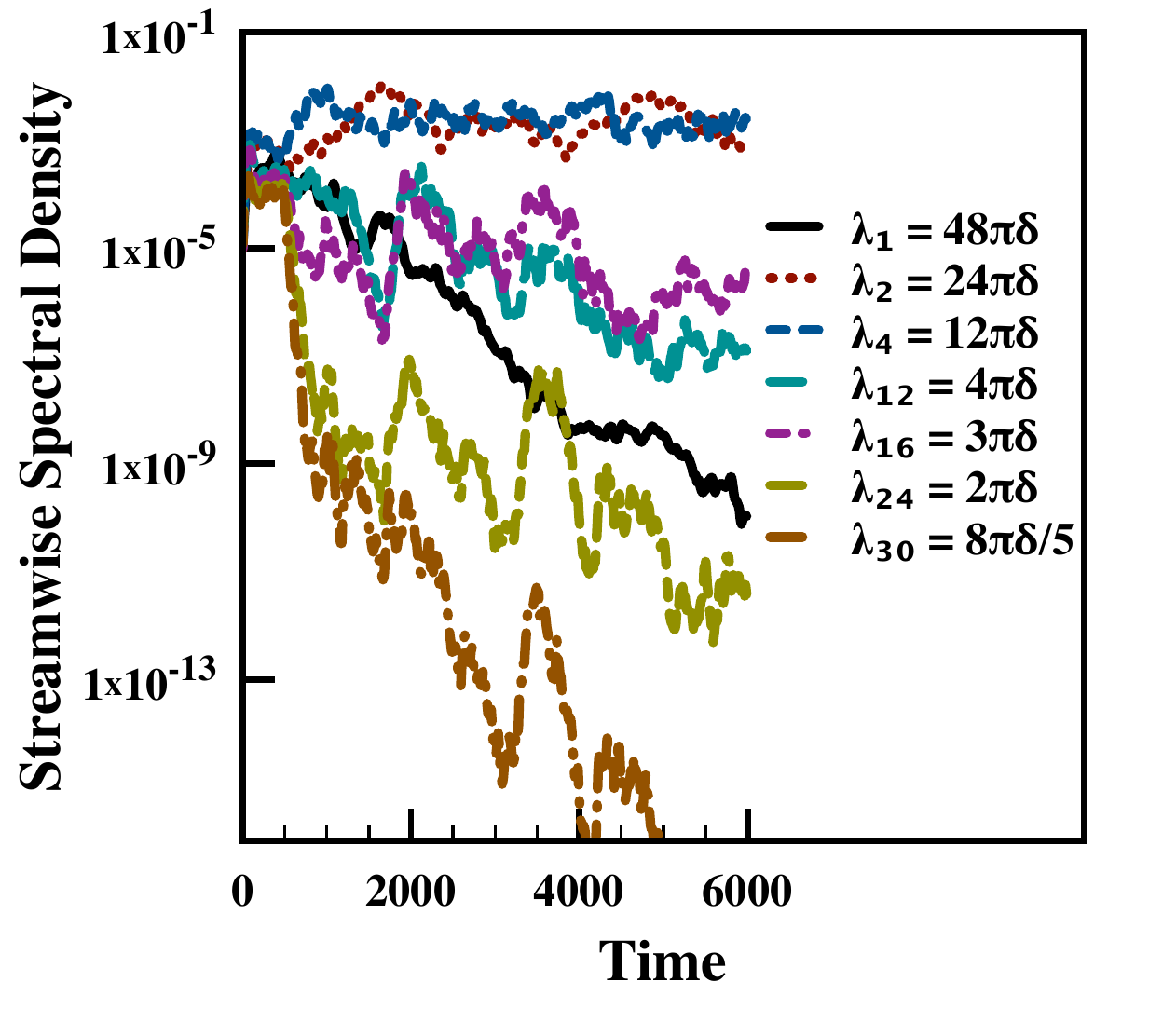} }
\caption{ Streamwise energy densities for (a) case RNL-F and (b) case RNL-G in Table \ref{table:geometry} at $R$ = 1000 with streamwise channel lengths $L_x=32\pi\delta$ and $L_x=48\pi\delta$ respectively.}
  \label{fig:rnlcollapses}
\end{figure}

Figures \ref{fig:32PRNLSpectra} and \ref{fig:48PRNLSpectra} respectively show the time evolution of the streamwise energy densities, $E_{\lambda_{n}}$, for cases RNL-F and RNL-G with respective streamwise channel lengths $L_x=32\pi\delta$ and $L_x=48\pi\delta$. In both of these simulations, turbulence is initiated by applying a stochastic excitation $\mathbf{e}$ in equation \eqref{eqn:RNL-perturb1}  from $t$ = 0 to $t$=500. Figure \ref{fig:32PRNLSpectra} reveals that in the channel with $L_x=32\pi\delta$ the streamwise energy density $E_{\lambda_{1}}$ is consistently lower than $E_{\lambda_2}$. It should be noted that this is the first channel length in this study where this occurs, i.e. in cases RNL-A, RNL-B, RNL-C and RNL-D, which are described in Table \ref{table:geometry}, the energy density associated with $\lambda_1$ has the greatest magnitude.  Figure \ref{fig:48PRNLSpectra} that for case  RNL-G the longest wavelength, $\lambda_1 = 48\pi\delta$, clearly decays to zero, confirming the existence of an upper wavelength limit of the natural support of RNL turbulence. The maximum wavelengths in the natural support of RNL turbulence in channels with streamwise extents of $L_x=48\pi\delta$ and $L_x=64\pi\delta$ are $\lambda = 24\pi\delta$ and $\lambda = 32\pi\delta$, respectively.  From the data across a number of channel configurations we estimate that the lower and upper bounds of the natural support for systems at $R$ = 1000  are approximately $2\pi\delta$ and $32\pi\delta$ respectively.

\begin{figure}[ht]
\includegraphics[width = 0.55\textwidth,clip=]{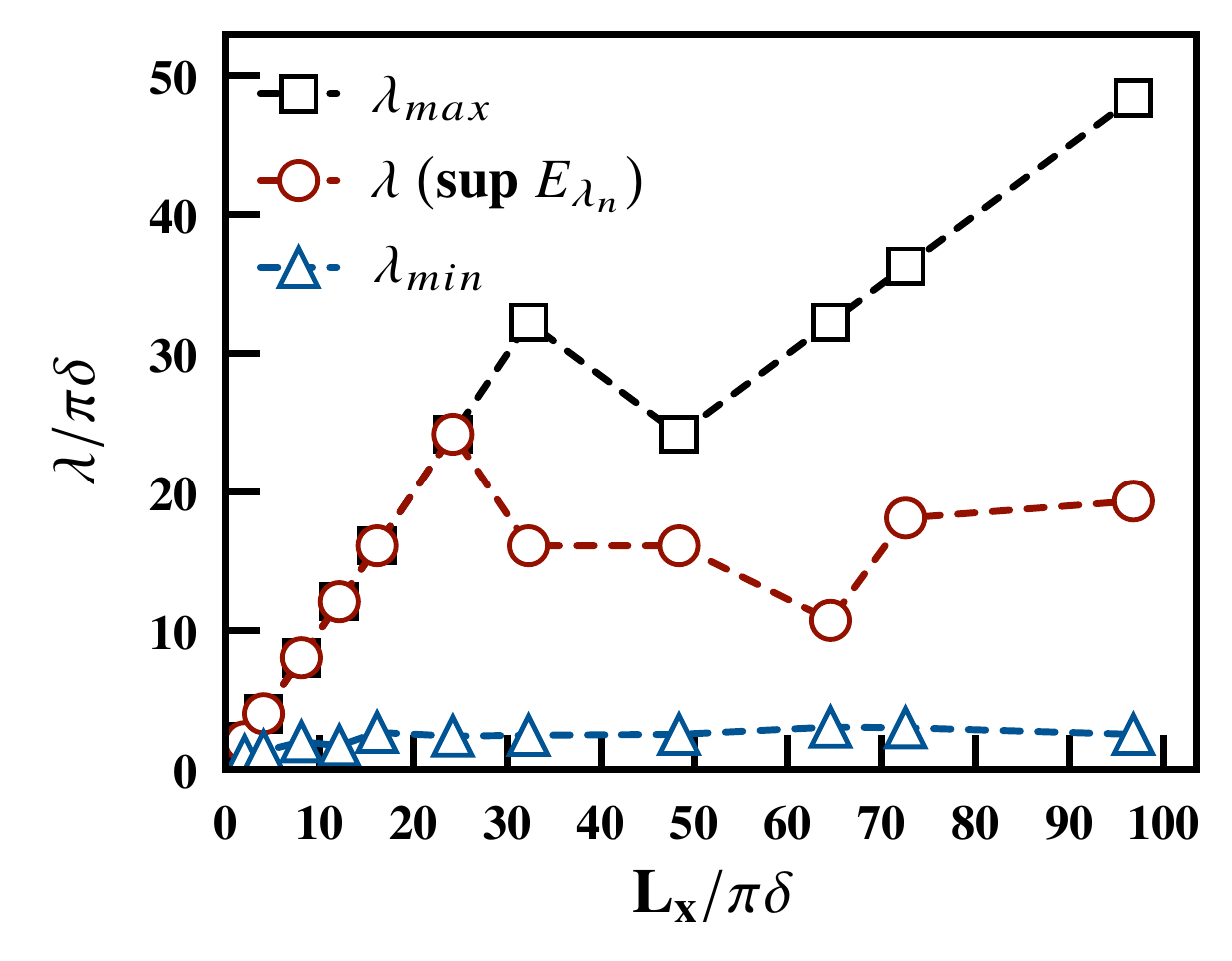}
\caption{(a) The longest wavelength, $\lambda_{max}$(black squares), shortest wavelength, $\lambda_{min}$ (blue triangles) and wavelength with the largest average energy density, $\lambda(\sup E_{\lambda_{n}})$ (red circles) in the natural set of the RNL system versus streamwise channel length, $L_x$, for $R$ = 1000. Lines are shown solely to guide the eye.}
  \label{fig:maxEnergyWavelength}
 \end{figure}
Figure \ref{fig:maxEnergyWavelength} shows the wavelength, $\lambda$, at which the largest time-averaged streamwise energy density occurs for each of the cases described in Table \ref{table:kxTruncationGeometry}. For RNL simulations with $L_x<32\pi\delta$ the longest wavelength has the highest mean energy density. However, for the RNL simulations in channels with $L_x$ $\ge$ 32$\pi\delta$, the most energetic wavelength is near 16$\pi\delta$. The most prominent  exception is the $L_x$ = 64$\pi\delta$ case. In this case the energy density for $\lambda_6 = 32\pi \delta/3$ is 1.7\% greater than the energy density associated with $\lambda_4 = 16\pi\delta$. We also note that cases RNL-E and RNL-G with respective channel lengths of 24$\pi \delta$ and 72$\pi \delta$ do not admit 16$\pi\delta$. In case RNL-E the two nearest wavelengths 24$\pi\delta$ and 12$\pi\delta$ have nearly identical energy densities, and for case RNF-G the maximum energy density occurs at $\lambda_4$ = 18$\pi\delta$. We conclude that 16$\pi\delta$ is a reasonable approximation for the ideal length for structures interacting with the mean flow regardless of the channel length.

Figure \ref{fig:maxEnergyWavelength} also shows the long and short wavelength limits, respectively denoted $\lambda_{max}$ and $\lambda_{min}$, of the natural set of the RNL system at $R$ = 1000. The short wavelength limit shows the same relative invariance with respect to the streamwise extent of the channel, $L_x$, as the wavelength of maximum energy density. For cases with  $L_x$ $\in$ $[16\pi,96\pi]\delta$ the short wavelength limit is constrained to a small range of values $\lambda$ $\in$ $[5\pi/2,3\pi]\delta$, whereas cases with $L_x<16\pi\delta$ have smaller $\lambda_{min}$ values. We do not see the same invariance in the long wavelength limit of the natural support. For simulations with $L_x$ $\in$ $[2\pi, 24\pi]\delta$ the long wavelength limit, $\lambda_{max}$, corresponds to $\lambda_1$ , the wavelength associated with the channel length. For channels with $L_x$ $\in$ $[32\pi, 96\pi]\delta$ the long wavelength limit, $\lambda_{max}$, corresponds to $\lambda_2$, the wavelength associated with the channel half-length. In describing the long wavelength limit, we are limited by the discretization in the streamwise direction. If there is an invariant for the long wavelength limit, it could only be discovered by examining cases with extremely large streamwise channel lengths; which is a direction for future work. For simulations in channels with $L_x$ $\in$ $[2\pi, 16\pi]\delta$ the largest time-averaged streamwise energy density, $\lambda(\sup E_{\lambda_{n}})$,  also occurs at wavelengths associated with the channel length, $\lambda_1$. For systems with $L_x$ $\in$ $[24\pi, 96\pi]\delta$ this maximum generally occurs at $\lambda$ = 16$\pi\delta$ unless the streamwise discretization does not permit this wavelength.  In those cases $\lambda(\sup E_{\lambda_{n}})$ occurs at the permitted wavelength that is closest to 16$\pi\delta$, i.e. for $L_x$ = 72$\pi\delta$ the maximum time-averaged streamwise energy density occurs at $\lambda_4$ = 18$\pi\delta$.

\subsection{Truncating the RNL support}
\label{sec:tuncatingRNLsupport}
The results in the previous section illustrate the natural support of the RNL system. We now show that RNL turbulence can be supported even in cases where the dynamics are restricted to a single streamwise varying perturbation interacting with the streamwise averaged mean flow. We then explore the relationship between the RNL system's natural support and the support of the truncated RNL system.  In Section \ref{sec:RNLnaturalsupport} we observed that simulations of case RNL-G exhibit both the upper and lower limit of the natural set, so we study the truncated systems in channels with $L_x$ = 48$\pi\delta$. The numerical details for the truncated RNL simulations are provided in Section \ref{sec:numerics} and a full list of the simulation parameters for these cases is given in Table \ref{table:kxTruncationGeometry}. For all of the truncated cases the simulations are initiated by applying the excitation $\mathbf{e}$ to \eqref{eqn:RNL-perturb1} for $t$ = 500 convective time units. We then apply a damping of $\xi$ in the manner shown in equation \eqref{eqn:RNL-perturb_trun} that varies from 0 to $1/\Delta t$ over a period of 100 convective time steps to all of the Fourier components in the perturbation equation (the streamwise varying $k_x\neq0$ modes) except for the single streamwise varying mode that we are limiting the dynamics to.  For the purpose of comparison we refer to the RNL system in which we do not damp any modes, i.e. $\xi=0$, as the baseline system, this is case RNL-G in Table \ref{table:geometry}.

\begin{figure}
\includegraphics[width = 0.5\textwidth,clip=]{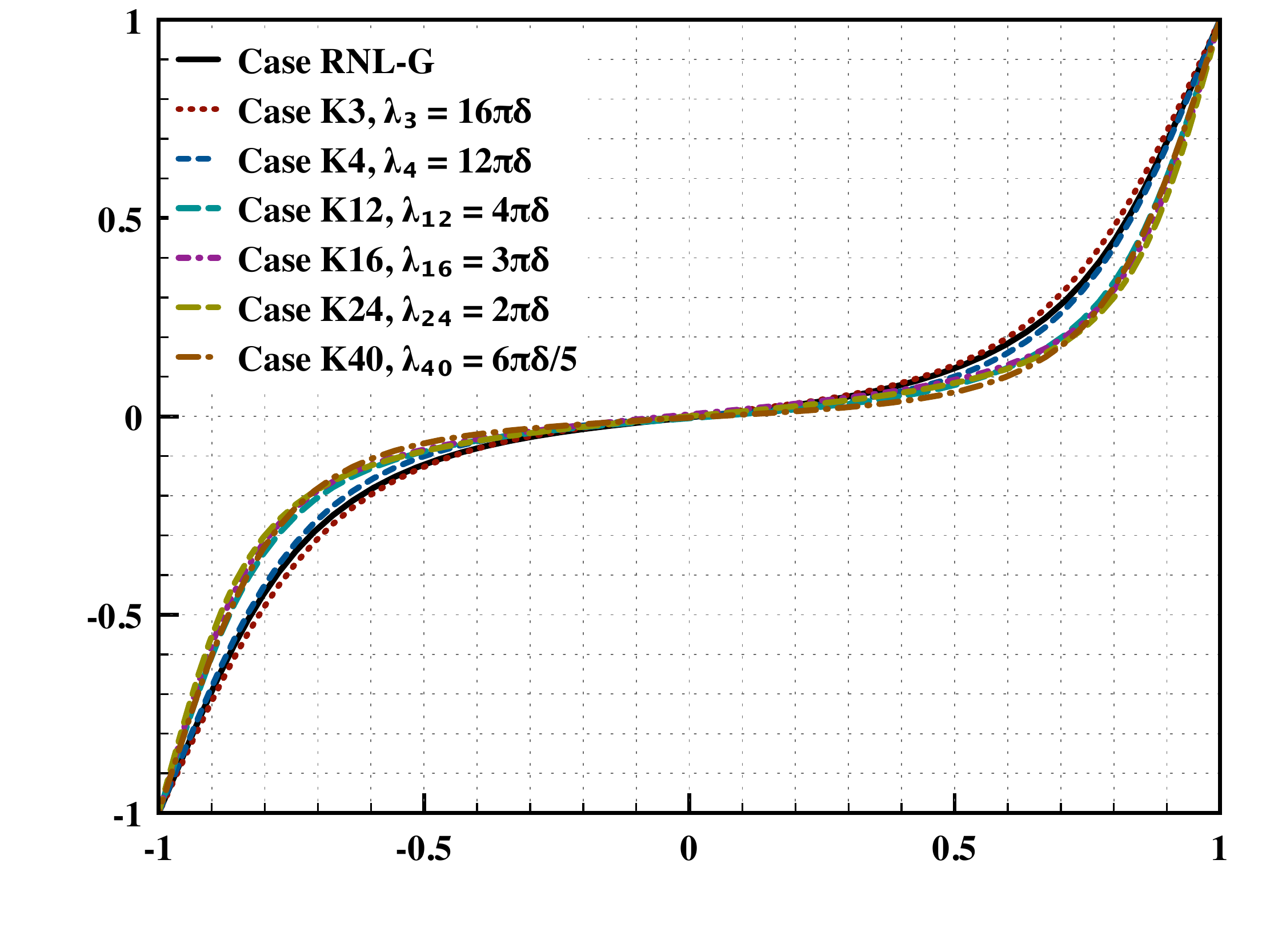}
\caption{Turbulent mean velocity profiles (based on a streamwise, spanwise and time average) obtained for cases RNL-G, K3, K4, K12, K16, K24, K40 all at $R=1000$ for a channel with $L_x$ =  48$\pi\delta$. Cases where the RNL is truncated to longer wavelengths exhibit a mean profile similar to the baseline RNL simulation (case RNL-G).}\label{fig:mean_singlekx}\end{figure}

Figure \ref{fig:mean_singlekx} demonstrates that truncated RNL systems all have an appropriately shaped mean velocity profile (based on a streamwise, spanwise and time average) and that these profiles are qualitatively similar to that obtained with the baseline RNL system.   In particular, for cases K3 and K4 the profiles are almost identical and have $R_\tau$ values of 53.4 and 57.0 respectively, which are close to the value of $R_\tau=56$ in the baseline system. Cases K12, K16, K24 and K40 show a higher shear stress at the wall and correspondingly higher $R_\tau$ values of 64.5, 69.6 and 64.5 respectively.
Selective filtering of streamwise varying perturbations in the RNL system was previously shown to strongly influence both the mean velocity profile and the spanwise spectra of the velocity field.\cite{Bretheim-etal-2014} Previous work has also shown that higher dissipation is associated with the inclusion of shorter wavelength perturbations.\cite{Thomas2014} This implies that the higher wavenumber and therefore higher dissipation cases K12, K16, K24 and K40 require a higher shear stress in order to attain a statistical equilibrium, which is the trend observed in the mean velocity profiles of Figure \ref{fig:mean_singlekx}.

\begin{figure}
\subfloat[]{\label{fig:kx3Spectra}
\includegraphics[width = 0.45\textwidth,clip=]{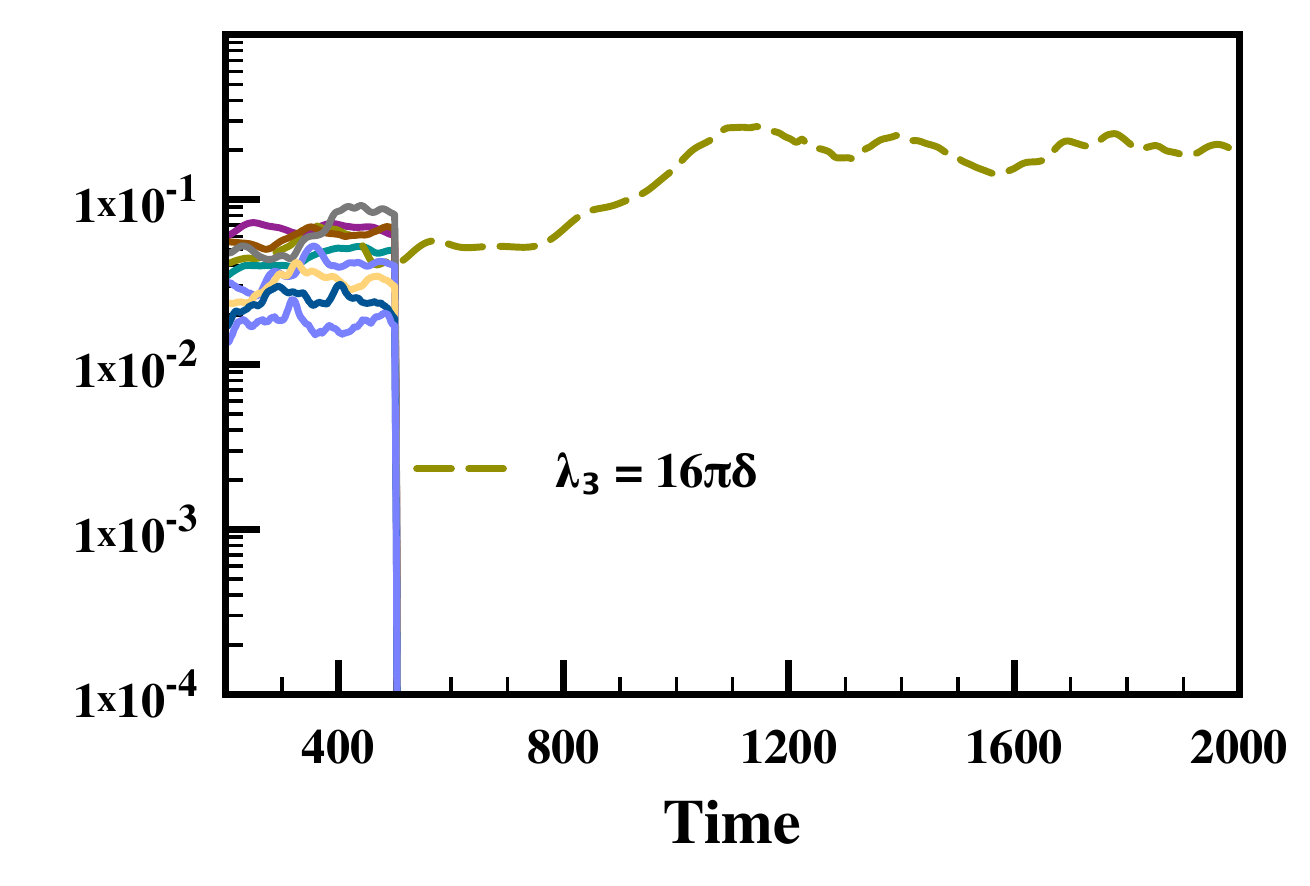}}
 \subfloat[]{\label{fig:kx4Spectra}
\includegraphics[width = 0.45\textwidth,clip=]{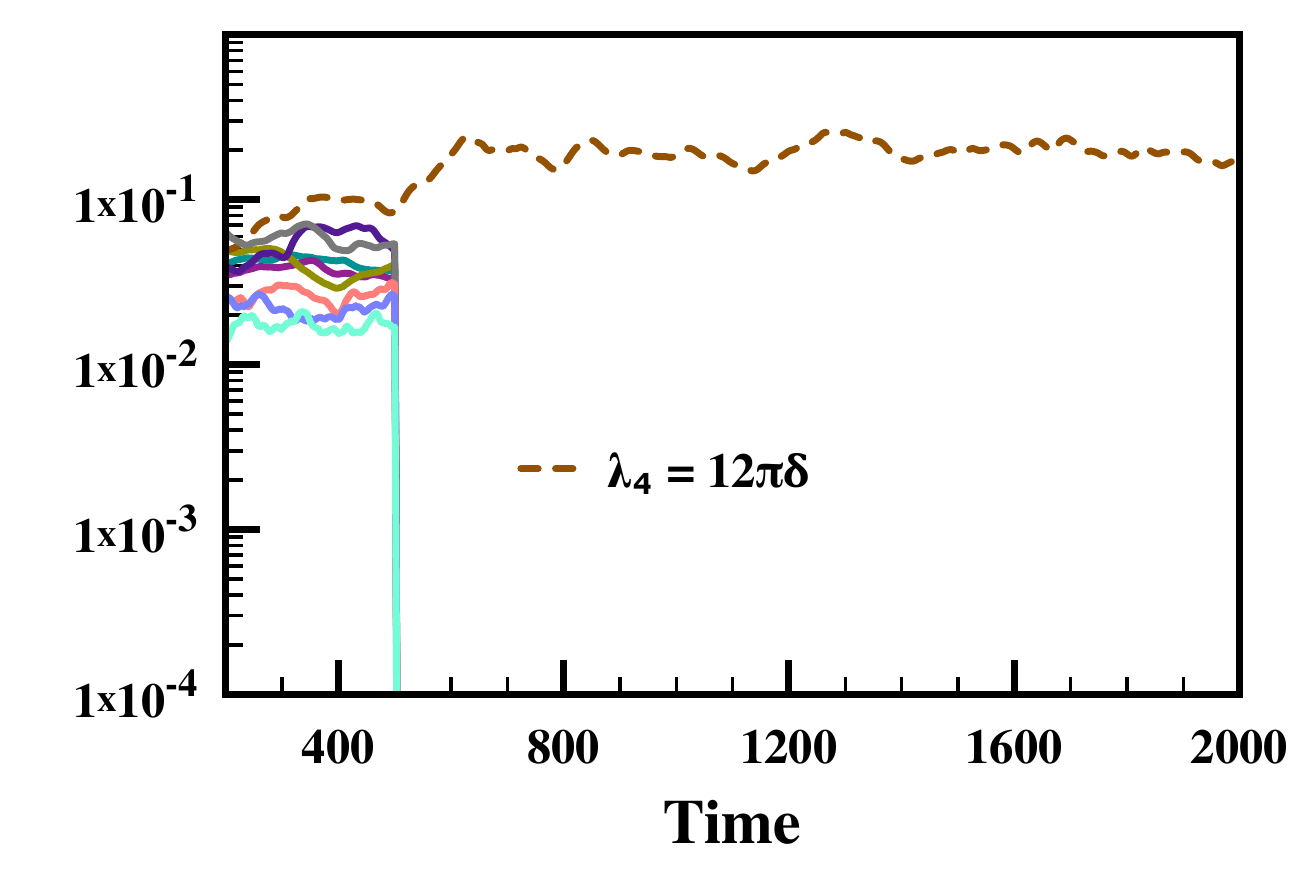}} \\
\subfloat[]{\label{fig:kx3RMS}
\includegraphics[width = 0.45\textwidth,clip=]{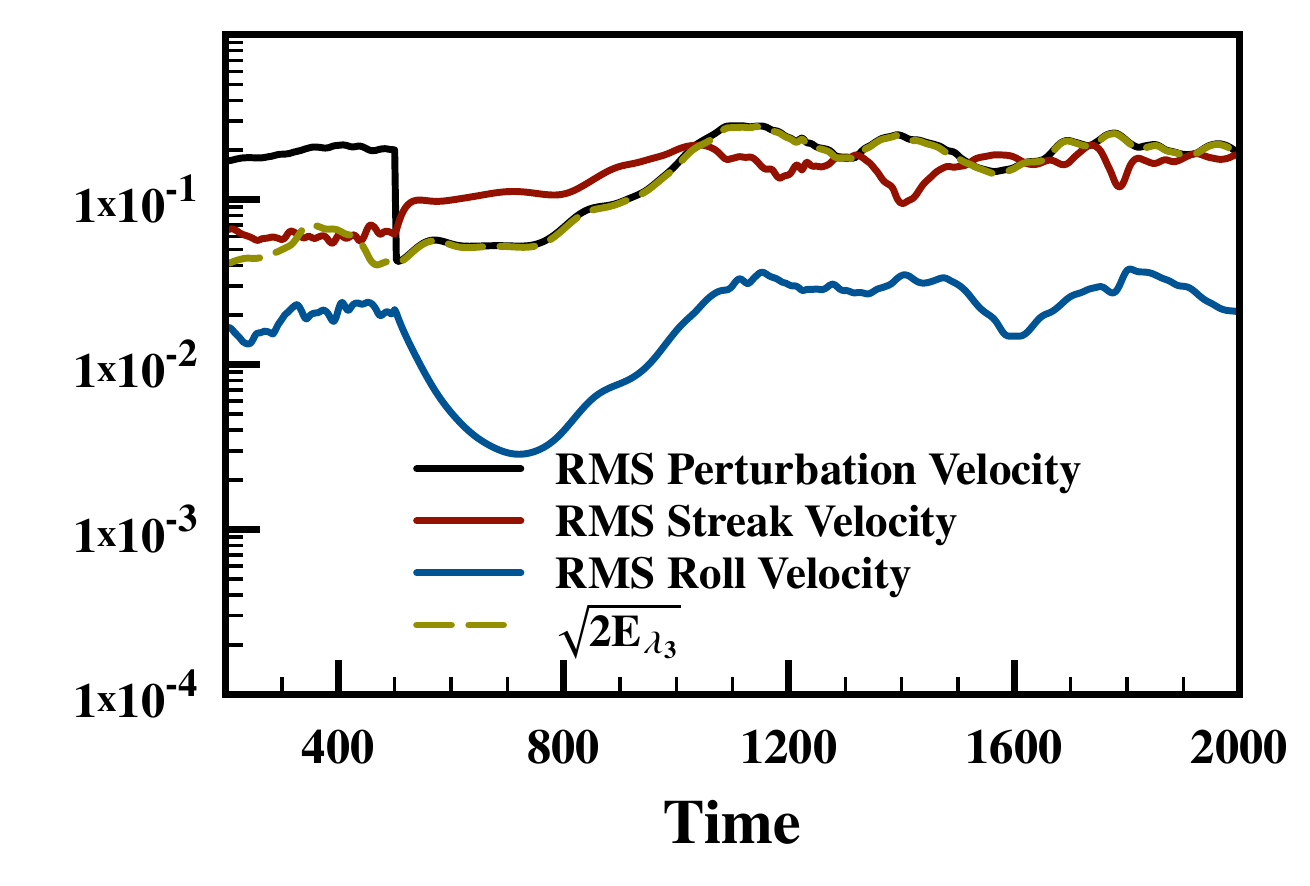}}
\subfloat[]{\label{fig:kx4RMS}
\includegraphics[width = 0.45\textwidth,clip=]{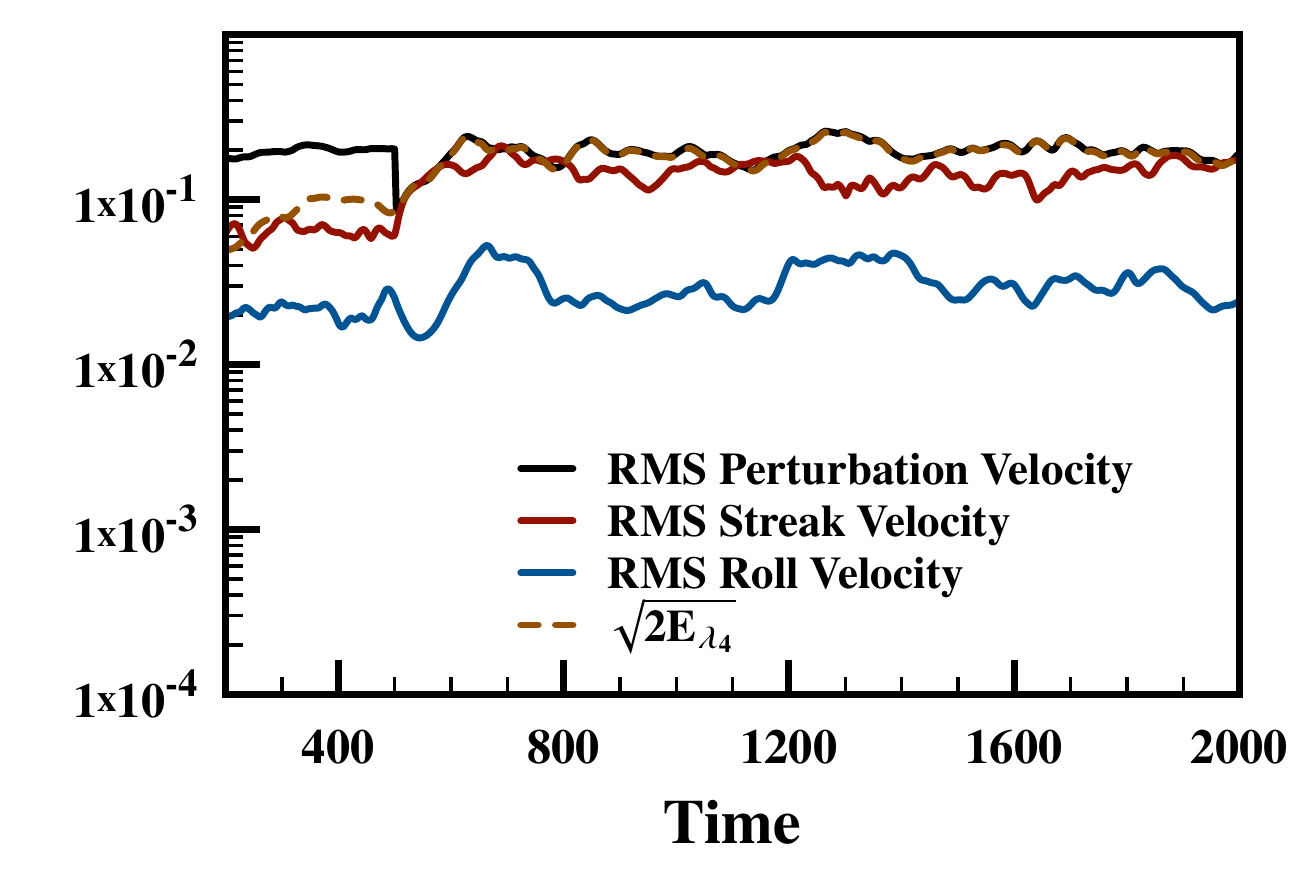} }
           \caption{Panels (a) and (b) show $\sqrt{2E_{\lambda_n}}$ versus time for cases K3 and K4, respectively. Panels (c) and (d) show $\sqrt{2E_{\lambda_n}}$ for the undamped wavelength ($\lambda_3=16\pi\delta$ and $\lambda_4=12\pi\delta$ respectively), the RMS perturbation velocity, $E_{\mathbf{u}}$, the RMS streak velocity, $U_s$, and the RMS roll velocity, $U_{Roll}$ for cases K3 and K4, respectively. In all panels, $R=1000$.}
\label{fig:kx3}
\end{figure}

Figures \ref{fig:kx3Spectra} and \ref{fig:kx4Spectra} show the time evolution of $\sqrt{2E_{\lambda_n}}$ for simulations of case K3 and case K4, respectively, in Table \ref{table:kxTruncationGeometry}. The wavelengths of the untruncated modes, respectively $\lambda_3$ = $16\pi\delta$ and $\lambda_4$ = $12\pi\delta$, are both within the natural support of the RNL model in a channel with $L_x$ = $48\pi\delta$ as shown in Figures \ref{fig:48PRNLSpectra} and \ref{fig:maxEnergyWavelength}. Both of these cases exhibit self-sustaining turbulent behavior despite the severe restriction in the streamwise harmonics. The rapid decay of the truncated wavenumbers subsequent to the removal of the stochastic excitation at $t$ = 500 is evident in figures \ref{fig:kx3Spectra} and \ref{fig:kx4Spectra}.

Figures \ref{fig:kx3RMS} and \ref{fig:kx4RMS} respectively replot $\sqrt{2E_{\lambda_n}}$ in figures \ref{fig:kx3Spectra} and \ref{fig:kx3Spectra} alongside the RMS perturbation velocity,
\begin{equation}
E_{\mathbf{u}}:=\sqrt{\sum_n 2 E_{\lambda_n}},
\label{eqn:pertub_rms}
\end{equation}
the RMS streak velocity,
\begin{equation}
U_s:=\sqrt{\int_0^{L_z} \int_{-\delta}^{\delta} \left(U -\left[U\right]\right)^2 \ dy \ dz},
\label{eqn:rms_streak}
\end{equation}
and the RMS roll velocity,
\begin{equation}
U_{Roll}:=\sqrt{\int_0^{L_z} \int_{-\delta}^{\delta} V^2+W^2 \ dy \ dz}.
\label{eqn:rms_roll}
\end{equation}
We note that in both simulations K3 and K4 both the RMS perturbation velocity, $E_{\mathbf{u}}$ and the RMS roll velocity, $U_{roll}$ return to levels comparable to those prior to the truncation, after a transient period. From figures \ref{fig:kx3RMS} and  \ref{fig:kx3RMS} it is clear that this adjustment occurs over a much longer time interval for case K3 as compared to case K4.

\begin{figure}[!t]
\subfloat[]{\label{fig:kx10Spectra}
\includegraphics[width = 0.45\textwidth,clip=]{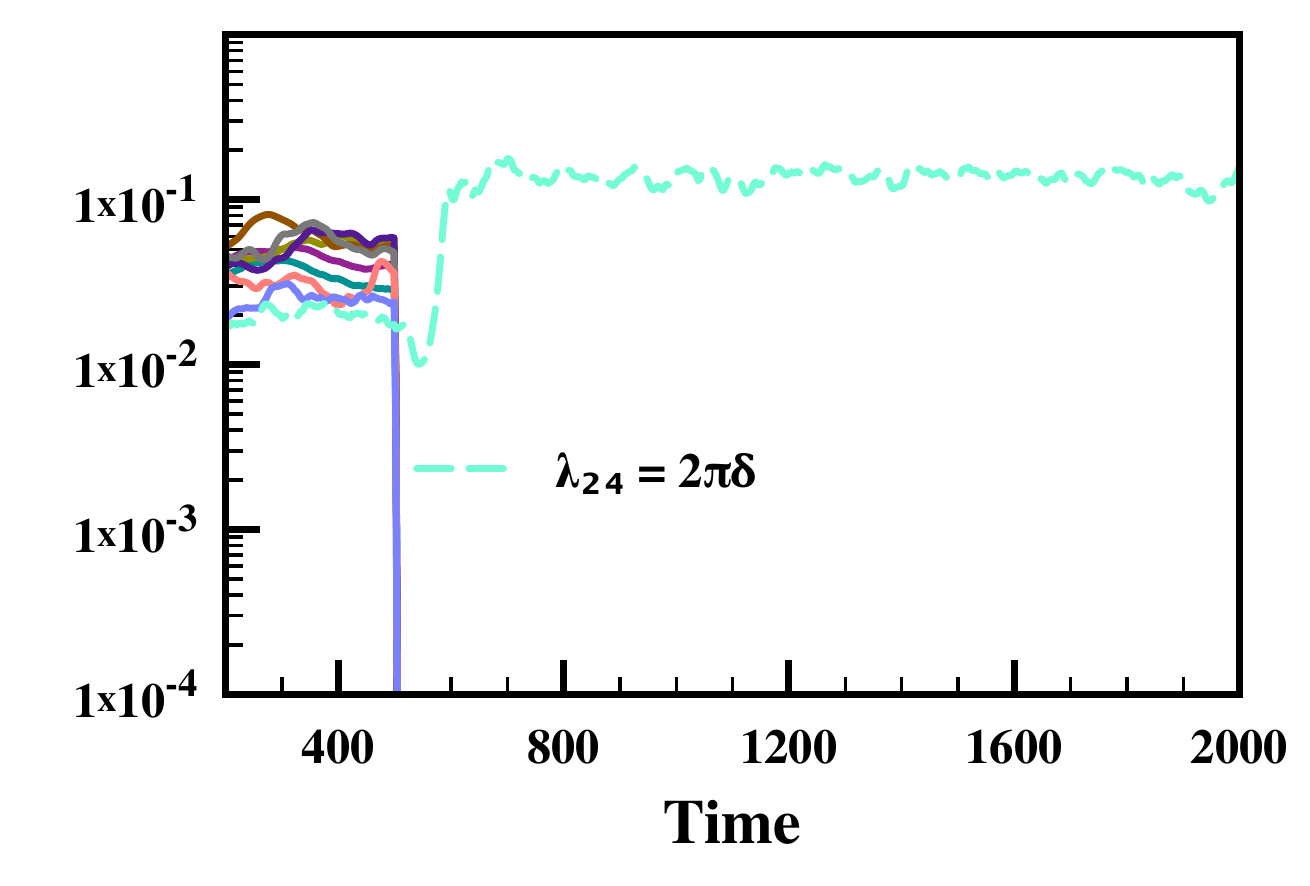}}
\subfloat[]{\label{fig:kx10RMS}
\includegraphics[width = 0.45\textwidth,clip=]{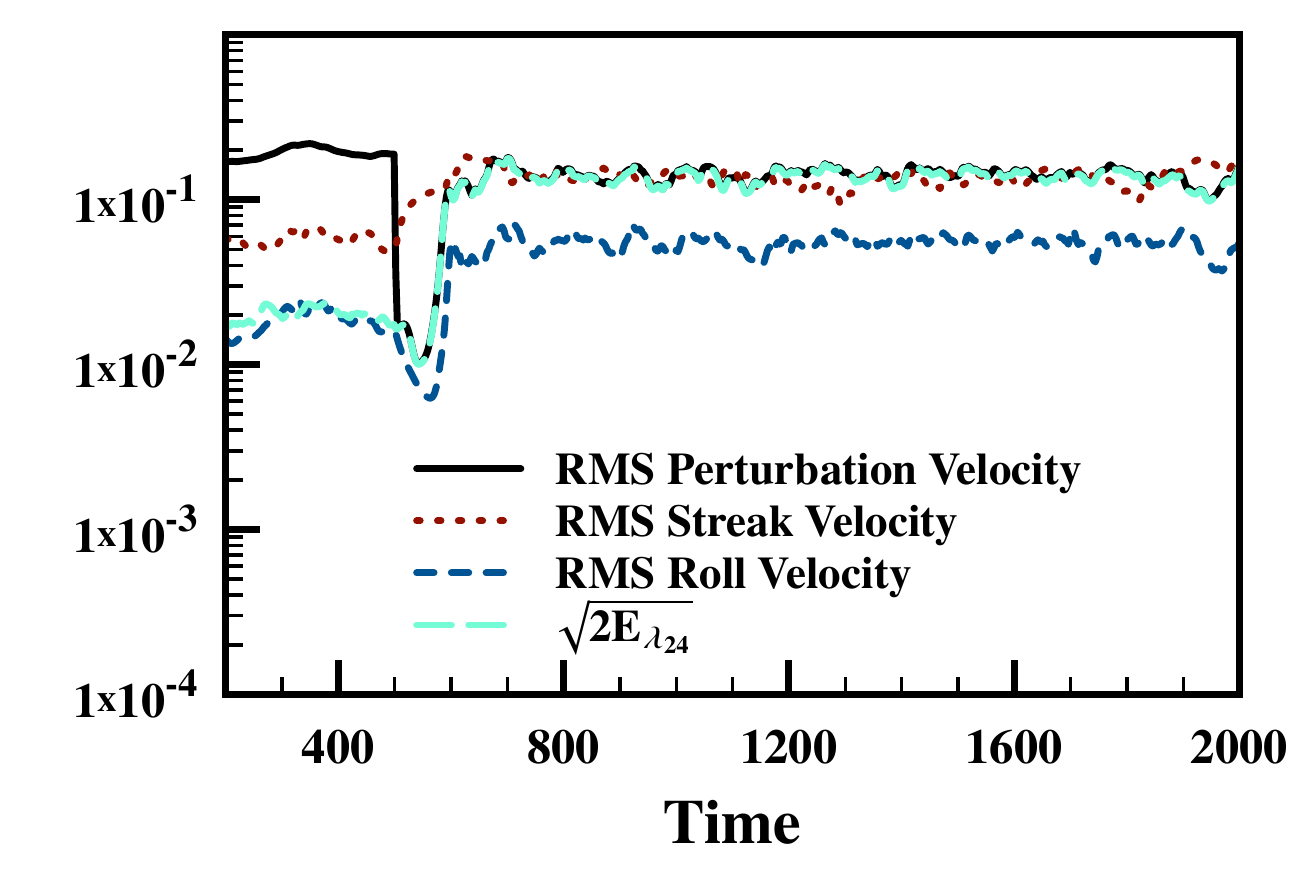} }
            \caption{(a) $\sqrt{2E_{\lambda_n}}$ versus time for case K24. (b) $\sqrt{2E_{\lambda_n}}$ for the undamped wavelength ($\lambda_{24}$), the RMS perturbation velocity, $E_{\mathbf{u}}$, the RMS streak velocity, $U_s$, and the RMS roll velocity, $U_{Roll}$ for case K24. In all panels, $R=1000$.}
            \label{fig:6}
\end{figure}

Figure \ref{fig:6} shows the same results as in figure \ref{fig:kx3} for case K24. The single undamped streamwise varying perturbation wavelength of $\lambda_{24}=2\pi\delta$ for this case is well outside of the natural set of the associated RNL-G (baseline) case, however the system is still able to sustain RNL turbulence with a reasonable mean profile as seen in Figure \ref{fig:mean_singlekx}. For this case $\sqrt{2E_{\lambda_{24}}}$ is approximately 5\% of the RMS perturbation velocity of the flow. When the damping is applied at $t$ = 500, the value of $\sqrt{2E_{\lambda_{24}}}$ increases by an order of magnitude such that $\sqrt{2E_{\lambda_{24}}}$ is nearly the same as the time-averaged RMS perturbation velocity prior to the application of the damping.

\begin{figure}
\subfloat[]{\label{fig:kx2Spectra}
\includegraphics[width = 0.45\textwidth,clip=]{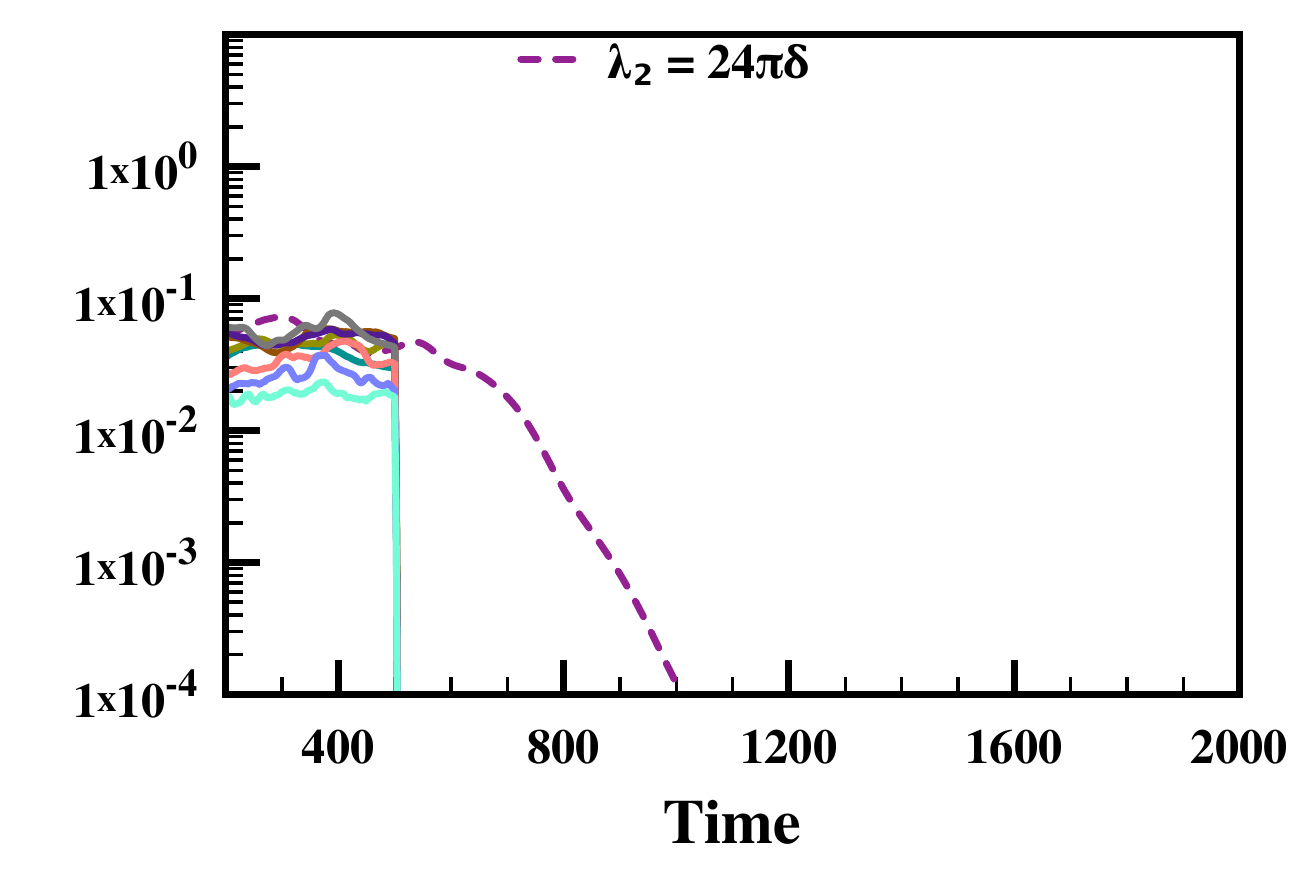}}
\subfloat[]{\label{fig:kx60Spectra}
\includegraphics[width = 0.45\textwidth,clip=]{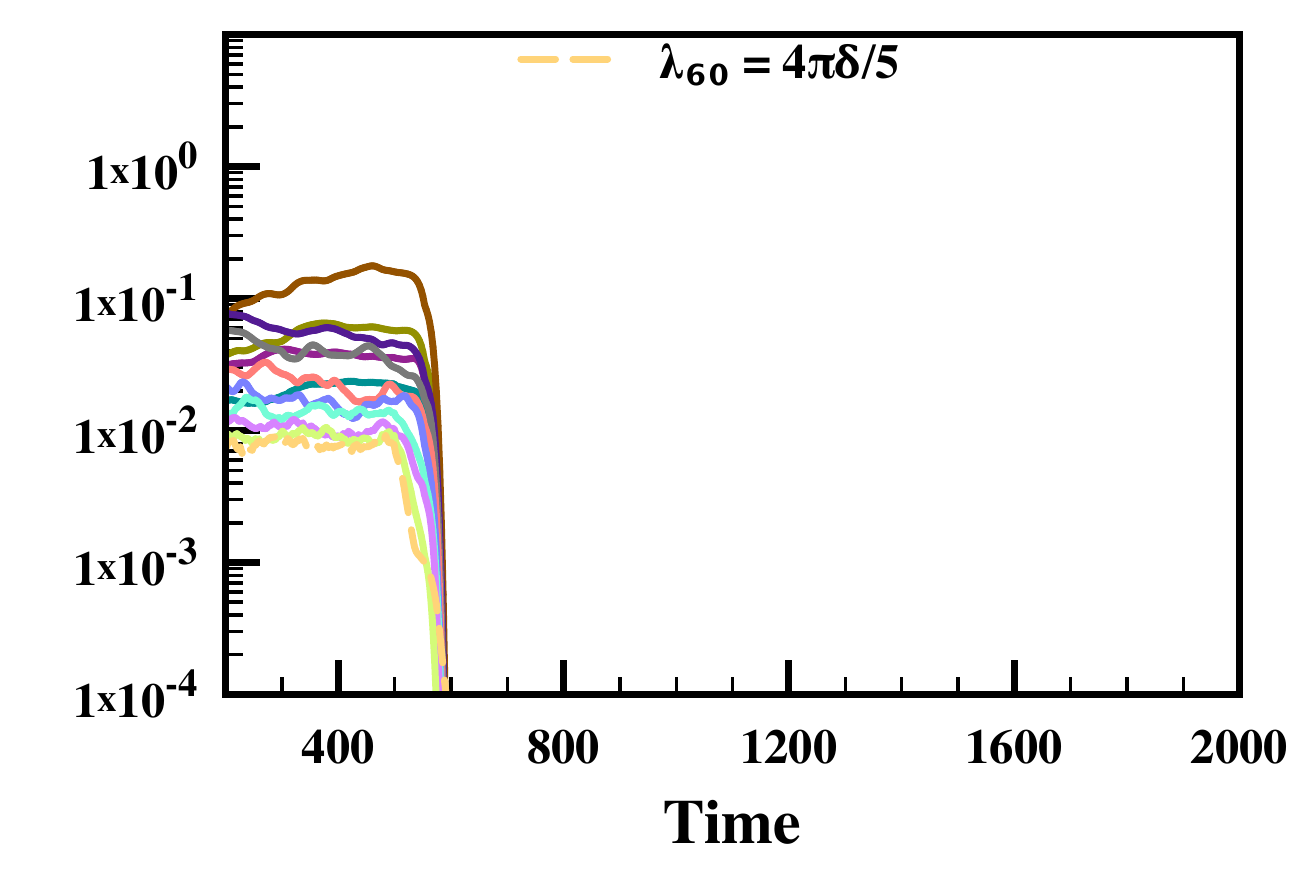} } \\
\subfloat[]{\label{fig:kx2RMS}
\includegraphics[width = 0.45\textwidth,clip=]{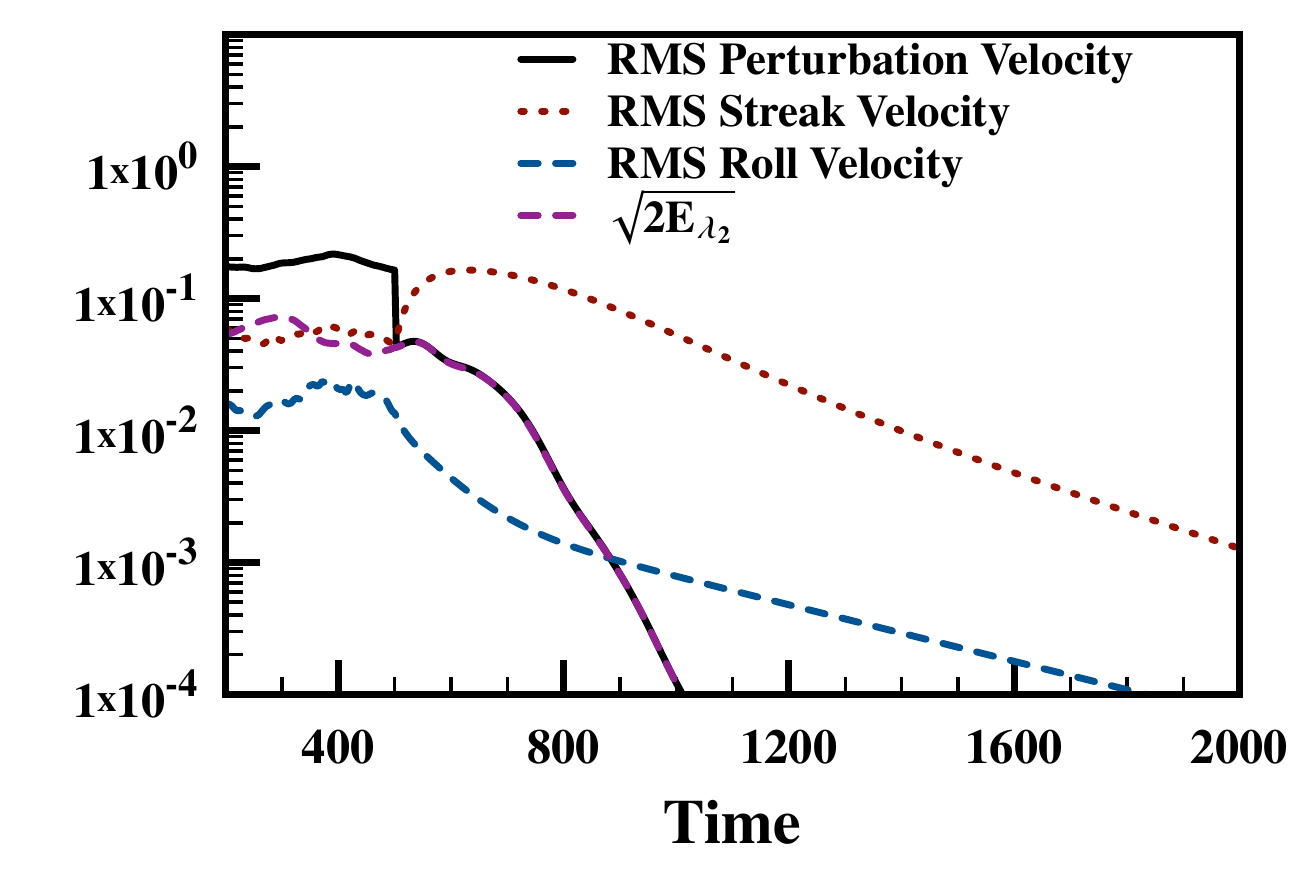}}
\subfloat[]{\label{fig:kx60RMS}
\includegraphics[width = 0.45\textwidth,clip=]{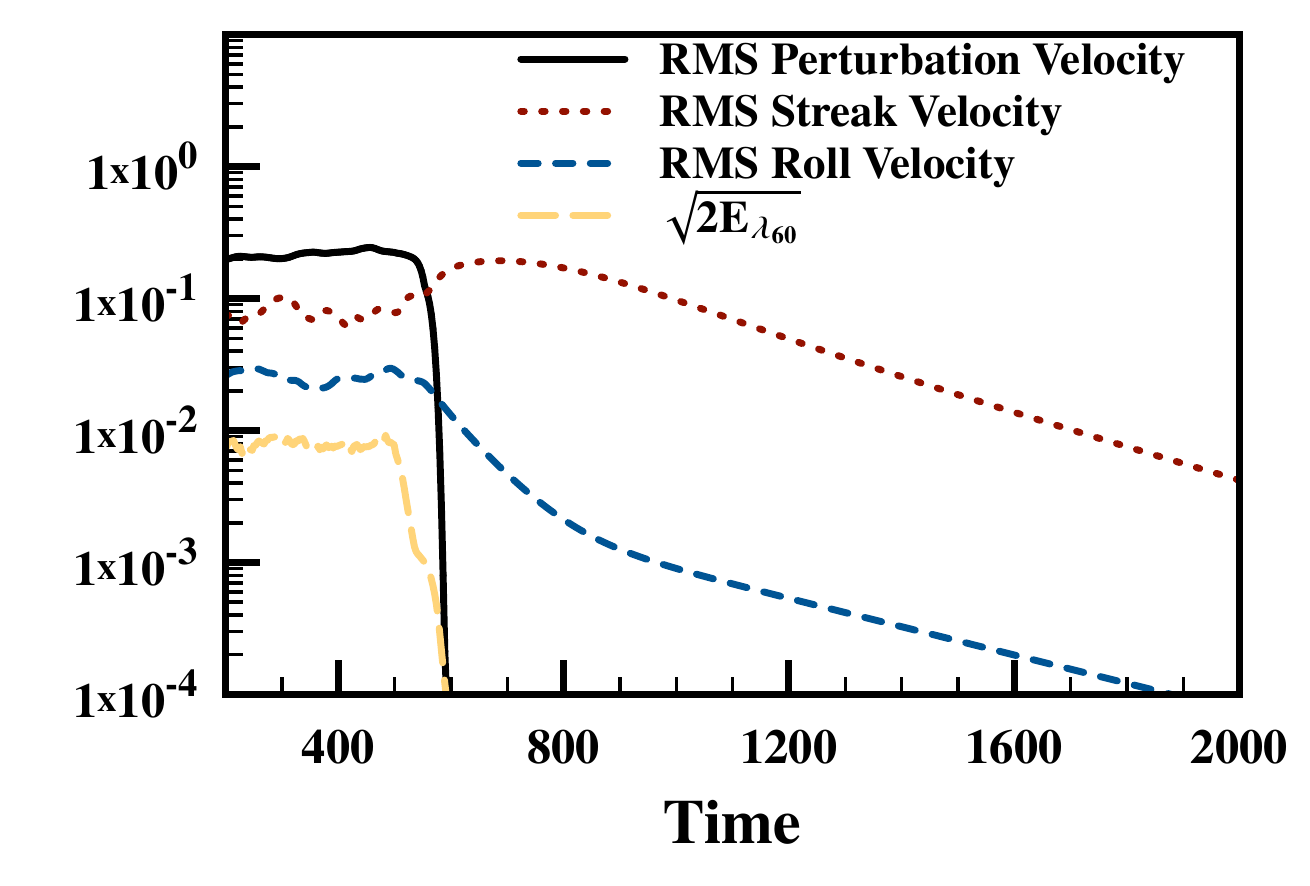} }
            \caption{Panels (a) and (b) show  $\sqrt{2E_{\lambda_n}}$ versus time resulting for cases K2 and K60, respectively. Panels (c) and (d) show $\sqrt{2E_{\lambda_n}}$ for the undamped wavelengths ($\lambda_{2}$ and $\lambda_{60}$ respectively), the RMS perturbation velocity, $E_{\mathbf{u}}$, the RMS streak velocity, $U_s$, and the RMS roll velocity, $U_{Roll}$ for case K2 and K60, respectively. In all panels, $R=1000$.}
\label{fig:kx2}
\end{figure}

Figure \ref{fig:kx2Spectra} shows the streamwise energy densities for case K2 in which only $\lambda_2$ = 24$\pi\delta$ and the mean flow are retained. In this case the flow becomes laminar after the damping is introduced and this behavior leads us to conclude that the RNL system in a $48\pi\delta$ channel cannot sustain turbulence when the dynamics are restricted to perturbation structures of length $\lambda_2=24\pi\delta$ interacting with the mean flow. The RNL system also cannot sustain turbulence for case K1 in which only $\lambda_1 = 48\pi\delta$ is retained.
\begin{figure}[ht]
\subfloat[]{\label{fig:48PitruncationSupport}
\includegraphics[width = 0.45\textwidth,clip=]{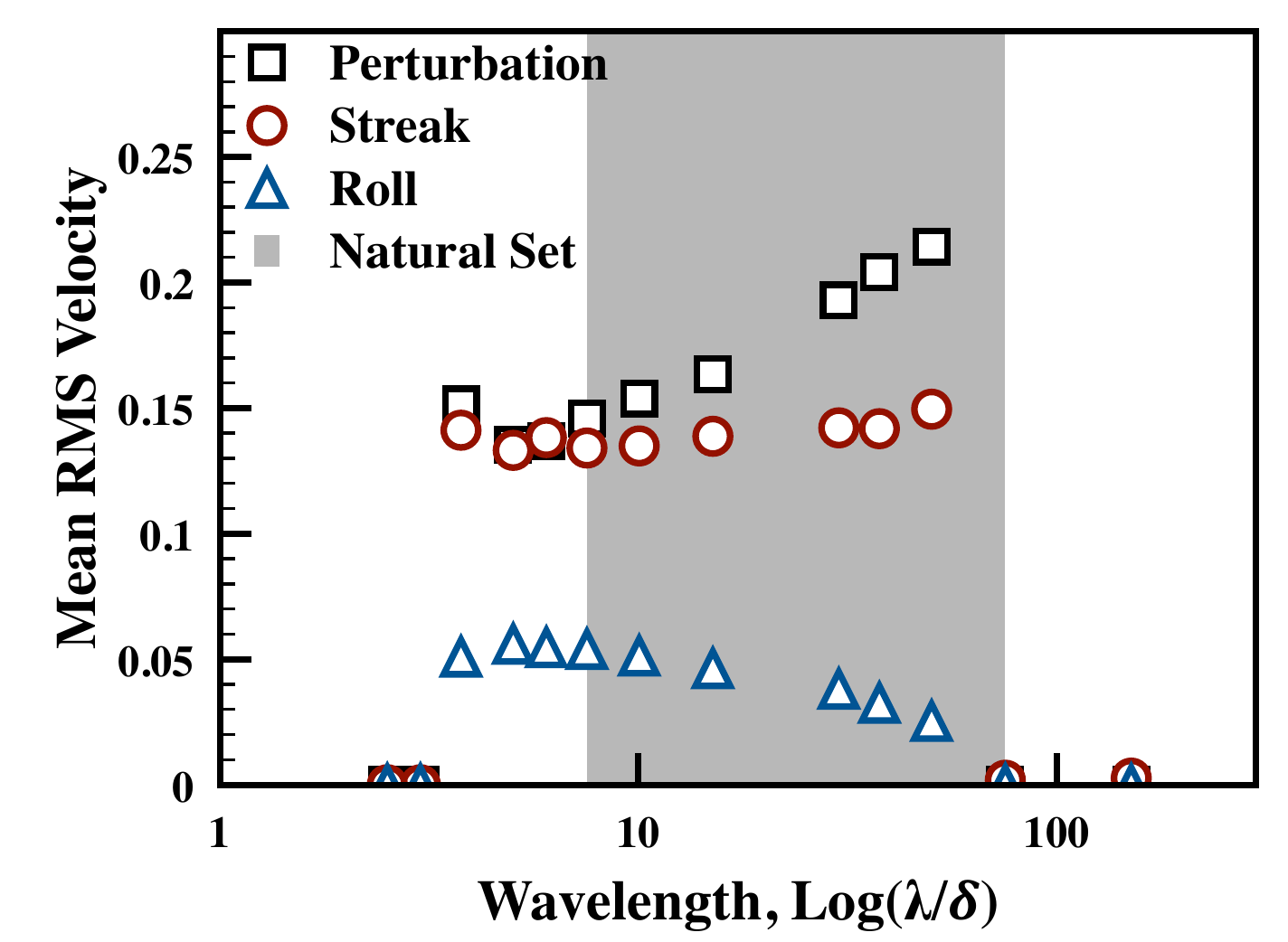}}
\subfloat[]{\label{fig4Pi:truncationSupport}
\includegraphics[width = 0.45\textwidth,clip=]{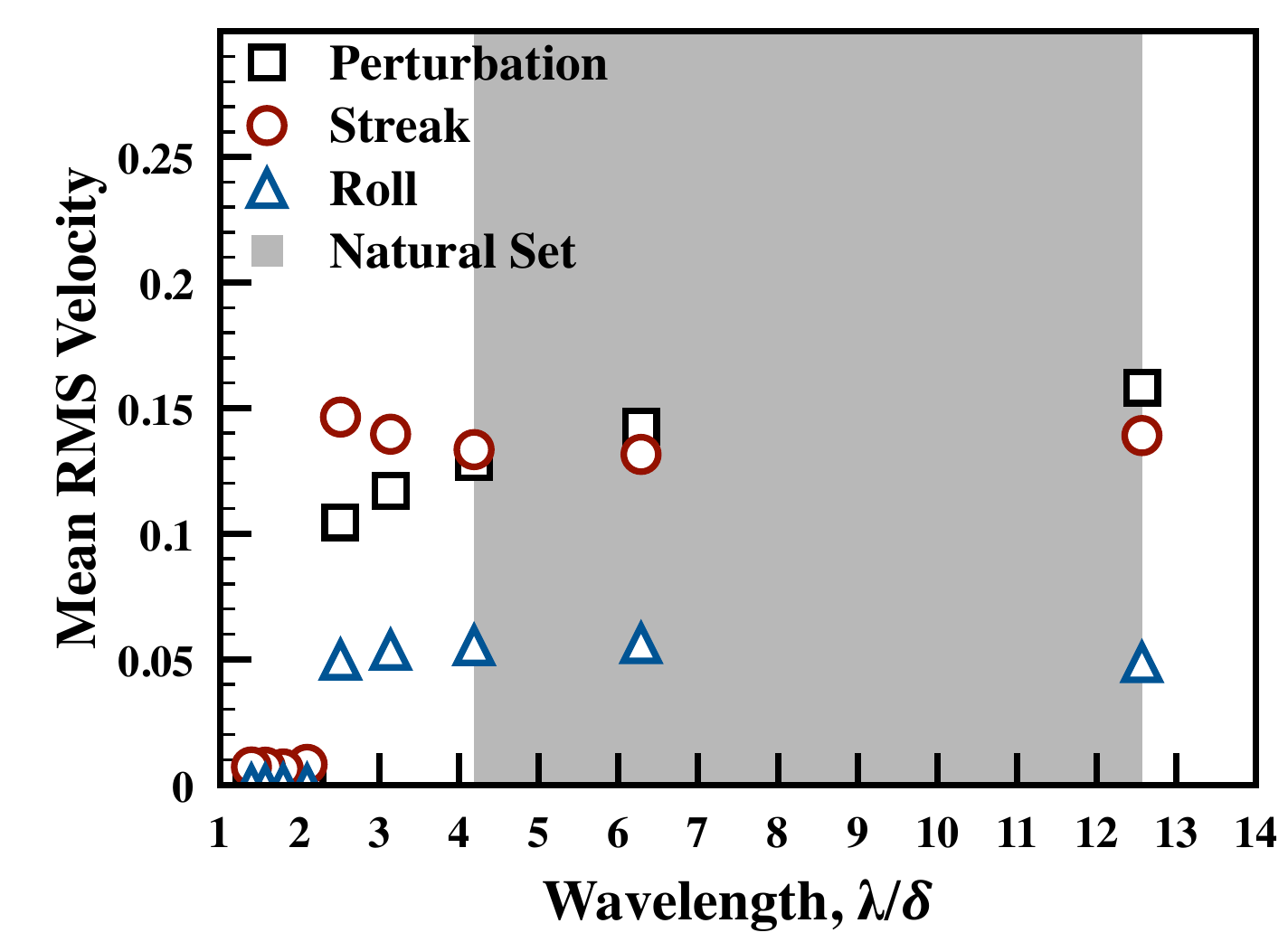}}
\caption{Time-averaged  RMS perturbation velocity, $E_{\mathbf{u}}$ (black squares), RMS streak velocity, $U_s$ (red circles), and RMS roll velocity, $U_{Roll}$ (blue triangles) for the truncated RNL systems in Table \ref{table:kxTruncationGeometry} for channels with (a) $\L_x$ = 48$\pi\delta$ and (b) $\L_x$ = 4$\pi\delta$. All simulations are at $R$ = 1000. The shaded region in each graph indicates the natural support for the RNL model in (a) a 48$\pi\delta$ channel and (b) a 4$\pi\delta$ channel. Both panels show that the set of perturbation wavelengths that sustain turbulence in the truncated RNL system is disjoint from the natural support of the RNL dynamics. The shortest wavelength at which simulations of the truncated RNL sustain turbulence is much smaller than the short wavelength limit of the natural support of the RNL system.}
  \label{fig:TruncationSupport}
 \end{figure}

As in the natural support of the RNL system, there is also a short wavelength limit beyond which turbulence is not robustly maintained in an RNL system that is truncated to a single streamwise varying mode interacting with the mean flow. This can be seen in figures \ref{fig:kx60Spectra} and \ref{fig:kx60RMS}, which show a short wavelength boundary for maintaining RNL turbulence with a single wavenumber. As in the case of the long wavelength boundary; the undamped wavelength, in this case $\lambda_{60}$, does not show any increase in energy. There is a region of uncertainty concerning the  maintenance of RNL turbulence in cases where a relatively short wavelength is retained. For the conditions used in K50, K60 and K70, some cases maintained turbulence while others relaminarized.  Similarly some of the simulations with the conditions of case D6, which corresponds to a truncated RNL simulation retaining only $\lambda_6$ = 2$\pi\delta$/3,   maintain turbulence and others relaminarize. However, in case D7 which retains only the mode associated with $\lambda_7$ = 4$\pi\delta$/7, the simulation always relaminarizes. Characterizing the factors that lead to the sustenance of turbulence versus relaminarization given the same simulation conditions is a topic of ongoing work.

A comparison between the natural support of streamwise varying modes that support RNL turbulence and the single streamwise varying modes that support turbulence in the truncated RNL system is provided in figures \ref{fig:48PitruncationSupport} and \ref{fig4Pi:truncationSupport}. For channels with $L_x= 48\pi\delta$, the wavelengths associated with the natural support extends from 2$\pi\delta$ to approximately 24$\pi\delta$. In contrast, the long wavelength limit associated with the truncated RNL system is approximately 16$\pi\delta$ and the short wavelength limit is well beyond the lower limit of the natural set and is closer to $\pi\delta$. A full characterization of the factors determining the bounds of the both the natural set of modes that support RNL turbulence and the set of single streamwise varying modes that can maintain turbulence through interactions with the mean flow will provide further insight into the nature of RNL SSP and are the subject of ongoing work.

\section{Conclusions and directions for future work}

In this paper we have shown that the RNL system intrinsically produces a minimal representation of turbulence that
 is supported by a streamwise averaged mean flow and a small set of streamwise varying perturbations.
 This minimal state arises spontaneously when the stochastic parametrization of the nonlinear interactions among the streamwise varying perturbations is eliminated in the perturbation equation.  The retained modes actively participate in the transfer of energy from the time-dependent streamwise averaged mean flow to the perturbation field and we refer to these as the natural support of self-sustaining turbulence in the RNL system.  We further show that the ability of the RNL system to self-sustain turbulence is remarkably robust. In particular, we demonstrate that RNL turbulence can be sustained when the dynamics are limited such that they comprise a single streamwise varying mode interacting with the streamwise averaged mean flow. The wavelengths of the streamwise varying perturbations that comprise the natural support of the RNL turbulence lie in a closed interval and our calculations suggest that this interval  becomes  independent of  channel size for long enough channels. The set of single streamwise wavenumbers that support RNL turbulence is found to extend to shorter wavelengths than those that are present in the natural set but not to longer ones.

The results presented here and the previously reported close correspondence between RNL simulations and DNS suggest that the fundamental mechanisms underlying wall-turbulence can be analyzed using highly simplified RNL systems.  The RNL framework provides distinct advantages over other minimal models that have been employed. First, the equations are directly derived from the NS equations and are easy to implement within an existing DNS code. Second, implementations of RNL minimal models do not rely on a particular Reynolds number or channel size and therefore Reynolds number trends as well as the dynamics of the RNL SSP in large channels can be explored using these models.  Finally, the RNL system inherently captures the dynamics of key flow structures as intrinsic elements of the system dynamics in a computationally and analytically tractable framework. These advantages make it a powerful tool for probing the dynamics of wall-turbulence. The insight gained through such studies can then be tested using DNS and exploited to develop flow control strategies.

\clearpage
\bibliography{S3T_turb_basic_refs}

\end{document}